\documentclass[useAMS,usenatbib]{mn2e}

\title[NGC~1866]
  {NGC~1866: a milestone for understanding the chemical evolution
   of stellar populations in the LMC
  \thanks{Based on observations collected at the ESO-VLT under program 074.D-0305.}}
   
\author[A. Mucciarelli et al.]
  {A.~Mucciarelli$^1$,
   S.~Cristallo$^2$,
   E.~Brocato$^3$,
   L.~Pasquini$^4$,
   O.~Straniero$^3$,
   \newauthor 
   E.~Caffau$^{5,6}$,
   G.~Raimondo$^3$,
   A.~Kaufer$^7$,
   I.~Musella$^8$,
   V.~Ripepi$^8$,
   M.~Romaniello$^4$,
   \newauthor 
   A.R.~Walker$^9$
   \\
  $^1$Dipartimento di Astronomia, Universit\`a 
  degli Studi di Bologna, via Ranzani 1, 40127
  Bologna, Italy
  \\
  $^2$Departamento de Fisica Teorica y del Cosmos, Universidad de Granada, 
  Campus de Fuentenueva, 18071, Granada, Spain  
   \\
   $^3$INAF - Osservatorio Astronomico di Collurania, via M. Maggini,
   64100, Teramo, Italy\\ 
   $^4$ESO - European Southern Observatory, Karl-Schwarzschild-str.2, 
   85748, Garching bei Munchen, Germany\\
   $^5$GEPI, Observatoire de Paris, CNRS, Universit\'e Paris Diderot, 
   92195, Meudon Cedex, France\\
   $^6$Zentrum f\"ur Astronomie der Universit\"at Heidelberg, Landessternwarte,
   K\"onigstuhl 12, 69117 Heidelberg, Germany\\
   $^7$ESO - European Southern Observatory, Alonso de Cordova 3107, Santiago, Chile \\
   $^8$INAF - Osservatorio Astronomico di Capodimonte, via di Moiariello 16, 80131 Napoli, Italy \\
   $^9$Cerro Tololo Inter-American Observatory, National Optical Astronomy Osbervatory, 
   Casilla 603, La Serena, Chile
    }

\usepackage{graphicx}

\pagerange{\pageref{firstpage}--\pageref{lastpage}} \pubyear{2002}

\def\LaTeX{L\kern-.36em\raise.3ex\hbox{a}\kern-.15em
    T\kern-.1667em\lower.7ex\hbox{E}\kern-.125emX}

\begin{document}

\label{firstpage}

\maketitle

\begin{abstract}

We present  new FLAMES@VLT spectroscopic observations of 30 stars in the field of the LMC stellar 
cluster NGC~1866.  NGC~1866 is one of the few young and massive globular 
cluster that is close enough so that its stars can be individually studied in detail.  
Radial velocities have been  used to separate stars belonging to the cluster and 
to the LMC field and the same spectra have been used to derive chemical abundances for a variety 
of elements, from [Fe/H] to the light (i.e. Na, O, Mg...) to the heavy ones. 
The average iron abundance of NGC~1866 turns out to be [Fe/H]=~--0.43$\pm$0.01 dex (with a dispersion 
$\sigma$=~0.04 dex), from the analysis of 14 cluster-member stars.
Within our uncertainties, the cluster stars are homogeneous, as far as chemical composition is concerned, 
independent of the evolutionary status. The observed cluster stars do not show any sign of the 
light elements 'anti-correlation' present in all the Galactic globular clusters so far studied, 
and also found in the old LMC stellar clusters. A similar lack of anti-correlations has been 
detected in the massive intermediate-age LMC clusters, indicating a different 
formation/evolution scenario for the LMC massive clusters younger than $\sim$ 3 Gyr with respect 
to the old ones.\\ 
Also opposite to the Galactic globulars, the chemical 
composition of the older RGB field stars and of the young post-MS cluster stars show robust homogeneity 
suggesting a quite similar process of chemical evolution. The field and cluster abundances are in 
agreement with recent chemical analysis of LMC stars, which show a distinctive chemical pattern for 
this galaxy with respect to the Milky Way. 
We discuss these findings in light of the theoretical scenario of chemical 
evolution of the LMC.
\end{abstract}


\section{Introduction}

The role of the Large Magellanic Cloud (LMC)  as an  exceptional laboratory for the study of 
stellar populations and stellar evolution has been early recognized by many authors 
\citep[e.g.][]{hdg60, hdg61, vdb68, vdbdb84}.
The star formation history and the related chemical evolution in the LMC have been studied through 
extensive photometric surveys \citep[see e.g.][]{harris09}  and theoretically through 
detailed modeling \citep{mb90}. The advent of the  8~m VLT telescopes  has
opened a new era in the investigation of resolved stellar populations,  
by producing high quality/high resolution spectra,  
which allow  the detailed chemical study of many single hot and cool stars in different regions 
of the LMC \citep[see e.g.][]{pomp}. One of the most distinctive results of these studies is that, 
similarly to other nearby dwarf galaxies, the LMC  shows clear signatures of a different 
chemical evolution with respect to the chemical evolution 
of the Milky Way sub-population components \citep{venn}.  

Another fundamental characteristic of the  LMC is that its cluster population covers a wide 
metallicity distribution and 
contains a large population of massive objects covering a 
wide age range, which provide a unique opportunity 
of studying rich samples of intermediate mass stars ($\sim$ 3-8 $M_{\odot}$) and the details of 
their evolutionary phases. A large and still ongoing effort has been done to collect 
photometric and spectroscopic data of stars in the stellar clusters of this galaxy 
\citep{hill00, pomp5, jo06, m08, m09, m10, tht09}.

In this scenario, NGC~1866 can be considered as a milestone for understanding 
the chemical evolution of the youngest stellar populations in the LMC, because this cluster 
is extremely rich ($\sim 5 \times 10^4$  $M_{\odot}$) compared with the coeval LMC clusters,  
with an age of $\sim 10^8 yr $ and mass of $\sim$ 5 $M_{\odot}$ 
for the stars evolving off the Main Sequence (MS) \citep{broc03} and a metallicity close to the 
one of 47 Tuc.
Concerning its metal content, the only study based on high-resolution 
spectra is that by \citet{hill00}, including Fe, O and Al abundances for three member stars 
of the cluster, providing an iron abundance of [Fe/H]=~--0.50$\pm$0.1 dex, a solar abundance 
of [O/Fe] and a mild depletion of [Al/Fe] with respect to the solar value.
 
Thus, high resolution spectroscopy properly coupled with a high quality color magnitude 
diagram (CMD) of NGC~1866 represent a unique tool to probe our knowledge of nucleosynthesis 
and mixing processes in intermediate mass stars during their evolution off of the MS.  
A further advantage of studying this cluster is that LMC  field stars can be easily identified 
as Red Giant Branch (RGB) stars, and a comparison between the abundances of these RGB field stars 
with those for the young cluster stars will be very powerful  to infer  
the chemical evolution processes in the LMC stellar population around the cluster and inside the 
cluster itself.  We take advantage of the large database of photometric data available 
for NGC~1866 and the related comparison with theoretical isochrones \citep{broc03}, and combine 
it with new high resolution spectra obtained at the VLT of stars well identified in the CMD of 
the LMC cluster NGC~1866 and its field. The paper is arranged as follows. 
The observations are described in the next section, while the assumptions on the stellar atmospheres 
are presented in Section 3. The chemical analysis and the related uncertainties are discussed in 
Section 4 and 5 and the results on the abundances of the elements are reported in section 6.  
Section 7 provides a general discussion on the observed framework, a brief summary concludes 
the paper.

\section{Observational material}

The spectroscopic data set analyzed here has been obtained with
the FLAMES spectrograph \citep{pasquini02} at VLT Kueyen 8.2~m
telescope, in the combined UVES+GIRAFFE mode, allowing the
simultaneous observation of 8 stars with the Red Arm of UVES at high-resolution
(R$\sim$42000)  and of 132 with GIRAFFE mid-resolution
(R$\sim$20000-25000) fibers.  All the observations have been
performed in Service Mode during 7 nights
between October 2004 and January 2005 under proposal 074.D-0305(A).\\
We used three different setups for the GIRAFFE observations:\\
(1) HR11 --- R=24200, $\Delta\lambda$=5597-5840 $\mathring{A}$; \\ 
(2) HR12 --- R=18700, $\Delta\lambda$=5821-6146 $\mathring{A}$;\\
(3) HR13 --- R=22500, $\Delta\lambda$=6120-6405 $\mathring{A}$. \\
The adopted GIRAFFE set-ups provide a spectral coverage
($\sim$5600-6400 $\mathring{A}$) including several absorption lines
of key elements such as iron, $\alpha$, iron-peak and neutron-capture
elements. All the targets have been observed in these three setups,
with a time exposure of 3600 sec for each individual exposure
(5 for HR11, 4 for HR12 and 3 for HR13), realizing a global S/N
ratio between 40 and 100 (per pixel) at $\sim$6000 $\mathring{A}$.
The spectra have been reduced by the standard
FLAMES reduction pipeline which includes bias subtraction,
flat-fielding, wavelength calibration with a reference
Th-Ar calibration lamp and final extraction of the 1-dimensional
spectra.

\begin{figure}
\includegraphics[width=85mm]{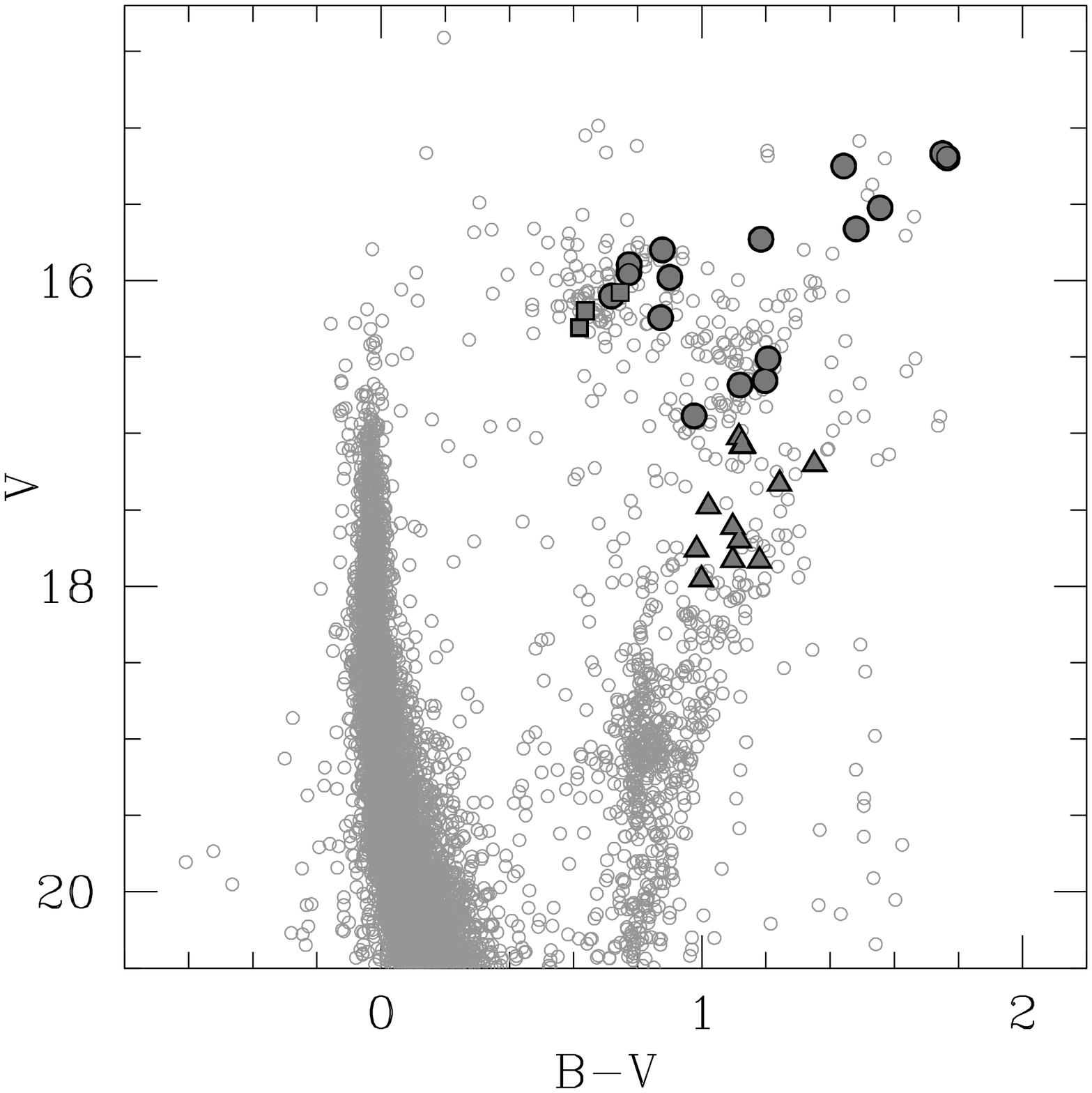}
\caption{Color-Magnitude Diagram of NGC~1866 \citep{musella06} with marked the observed target stars:
the cluster-member stars and the LMC field stars analyzed in this work are marked 
respectively as filled circles and triangles. Squares indicate the observed Cepheid stars.}
\label{cmd}
\end{figure}

The radial velocity of each spectrum has been derived with the
cross-correlation task of the BLDRS (GIRAFFE Base-Line Data
Reduction Software \footnote{http://girbldrs.sourceforge.net/}),
while for the stars observed with UVES the radial velocity has
been estimated by measuring the centroids of several tens of
un-blended lines. Heliocentric corrections have been computed by
using the IRAF task RVCORRECT.  The stars with $v_{helio}<$200 km
$s^{-1}$ have been discarded because they likely belong to our Galaxy,
according to the radial velocity maps computed for the LMC by
\citet{stav}. We obtained an average heliocentric velocity for the
cluster of $v_{helio}$=298.5$\pm$0.4 km $s^{-1}$ ($\sigma$=1.6 km
$s^{-1}$) by using 16 stars, in good agreement with the previous
determination by \citet{hill00} of $v_{helio}$=299.8$\pm$0.5 km
$s^{-1}$ ($\sigma$=1.4 km $s^{-1}$). In the computation of the
average radial velocity we have excluded three observed Cepheid
stars. Moreover, 11 RGB stars belonging to the LMC
field have been observed, with  $v_{helio}$ ranging from 261.4 to 305.5
km $s^{-1}$.
All the individual exposures have been sky-subtracted, shifted to
zero-velocity, then co-added and normalized to unity.
Fig.~\ref{cmd} shows the CMD in the
V-(B-V) plane of NGC~1866 with the positions of our
target stars:
big grey circles indicate the stars member of NGC~1866 (according
to their $v_{helio}$ value, distance and position in the CMD),
grey triangles are the observed LMC field stars and grey squares the 
Cepheids. Information about
all observed targets is listed in Tab. \ref{info} with ID number 
\citep{musella06}, RA, Dec,
the V and K magnitudes, heliocentric radial velocities and S/N ratio
(computed at $\sim$6000 $\mathring{A}$). The total sample consists of 30 stars, 
of which 19 are from the cluster and 11 from the LMC field. 
The three cluster Cepheids will  be discussed in a forthcoming paper.

\begin{table*}
\begin{center}
\begin{minipage}{150mm}
\caption{Target information: ID number, RA, Dec, V and K magnitudes, heliocentric radial velocities, S/N 
and membership.} 
\label{anymode}
\begin{tabular}{lcccccccc}
\hline
ID-Star   & RA & Dec &  V &    K & $v_{helio}$  & S/N & Membership & Notes \\
    & (J2000)  &  (J2000)  & & & (km$s^{-1})$ & & & \\
\hline    
   652  &  78.384167 &  --65.509056   & 17.76  &  15.02 & 292.9  & 40    & FIELD  &   \\
  1025  &  78.342208 &  --65.503500   & 16.20  &  14.60 & 294.9  & 80    & CLUSTER  &  Cepheid --- HV12197 \\
  1146  &  78.366417 &  --65.501639   & 15.20  &  10.94 & 299.0  & 120   & CLUSTER  & UVES --- TiO bands  \\
  1491  &  78.450708 &  --65.497028   & 17.95  &  15.46 & 267.3  & 45    & FIELD  &    \\
  1605  &  78.282292 &  --65.495417   & 17.33  &  14.37 & 266.7  & 45    & FIELD  &    \\
  1969  &  78.354708 &  --65.491444   & 16.31  &  14.66 & 311.0  & 90    & CLUSTER  &  Cepheid --- HV12199   \\
  1995  &  78.533833 &  --65.491222   & 17.08  &  14.40 & 280.3  & 50    & FIELD  &    \\
  2131  &  78.449917 &  --65.489694   & 15.66  &  12.25 & 299.1  & 100   & CLUSTER  &     \\
  2305  &  78.357125 &  --65.487639   & 17.61  &  14.79 & 272.2  & 45    & FIELD  &    \\
  2981  &  78.403542 &  --65.481611   & 15.52  &  11.95 & 301.3  & 100   & CLUSTER  & UVES   \\
  4017  &  78.334708 &  --65.474111   & 16.51  &  13.72 & 298.7  &  70   & CLUSTER  &    \\
  4209  &  78.347917 &  --65.472972   & 17.20  &  13.96 & 270.8  &  60   & FIELD  &   \\
  4425  &  78.374708 &  --65.471500   & 15.73  &  12.98 & 299.3  &  90   & CLUSTER  &    \\
  4462  &  78.497500 &  --65.471333   & 15.80  &  13.78 & 298.8  &  80   & CLUSTER  &   \\
  5231  &  78.411667 &  --65.466500   & 15.24  &  11.86 & 298.1  & 100   & CLUSTER  &     \\
  5415  &  78.435583 &  --65.465194   & 15.90  &  14.02 & 297.6  &  90   & CLUSTER  &     \\
  5579  &  78.421167 &  --65.464028   & 16.09  &  13.94 & 291.7  &  90   & CLUSTER  &  Cepheid --- We2   \\
  5706  &  78.454875 &  --65.463028   & 16.65  &  13.83 & 298.5  &  80   & CLUSTER  &   \\
  5789  &  78.413625 &  --65.462389   & 15.97  &  13.80 & 297.2  &  90   & CLUSTER  &     \\
  5834  &  78.443333 &  --65.462056   & 15.17  &  10.78 & 296.0  & 120   & CLUSTER  & UVES --- TiO bands \\
  7111  &  78.476333 &  --65.451861   & 17.83  &  15.11 & 261.4  &  40   &  FIELD  &    \\
  7392  &  78.422375 &  --65.449361   & 15.95  &  14.06 & 297.9  &  85   & CLUSTER  &   \\
  7402  &  78.361208 &  --65.449250   & 16.88  &  14.53 & 297.8  &  60   & CLUSTER  &    \\
  7415  &  78.433625 &  --65.449167   & 16.24  &  14.14 & 302.2  &  70   & CLUSTER  &    \\
  7862  &  78.458417 &  --65.444750   & 16.68  &  13.99 & 297.2  &  60   & CLUSTER  &    \\
  9256  &  78.489750 &  --65.428778   & 17.48  &  15.09 & 293.4  &  40   &  FIELD  &    \\
  9649  &  78.509167 &  --65.424056   & 17.02  &  14.43 & 272.1  &  60   &  FIELD  &    \\
 10144  &  78.482625 &  --65.415944   & 17.83  &  14.90 & 273.2  &  50   &  FIELD  &    \\
 10222  &  78.530208 &  --65.414583   & 17.70  &  14.81 & 305.5  &  40   &  FIELD  &   \\
 10366  &  78.430875 &  --65.412111   & 16.10  &  14.36 & 296.7  &  60   & CLUSTER  &   \\
\hline
\label{info}
\end{tabular}
\end{minipage}
\end{center}
\end{table*}

\section{Atmospheric parameters}

Initial atmospheric parameters have been computed from the photometric data. 
Effective temperatures ($T_{eff}$) for the target stars have been
derived from de-reddened (V-K) color, obtained by combining the
visual FORS1 photometry 
(\citet{musella06}, Musella et al. 2010, in preparation)
and the near-infrared SOFI photometry
\citep{m06a}. We assumed a reddening value of E(B-V)~=0.064 by
\citet{walker}, the extinction law by \citet{rl85} and using the
empirical $(V-K)_0$-$T_{eff}$ calibration computed by
\citet{alonso} and based on the Infrared Flux Method;
transformations between the different photometric systems have
been performed by means of the relations by \citet{carpenter} and
\citet{alonso98}.

Surface gravities have been obtained from the classical equation
$$log~(g/g_{\odot}) = 4\cdot log(T_{eff} / T_{eff,\odot})$$ 
$$+ log~(M/M_{\odot}) - 0.4\cdot (M_{bol}-M_{bol,\odot})$$
by adopting a distance modulus of $(m-M)_0$=~18.50, the bolometric corrections
computed by \citet{alonso}. We consider a mass of $M_{1866}$=~4.5 $M_{\odot}$
\citep[according to the cluster age inferred by][]{broc03}
for the cluster-member stars and of $M_{LMC-Field}$=~1.5
$M_{\odot}$ (corresponding to the typical evolutive mass of a
population of $\sim$2 Gyr) for the LMC field stars.
We checked that photometric $T_{eff}$ and log~g well
satisfy the excitation and ionization equilibrium, respectively; hence
the neutral iron abundance must be independent by the excitation
potential $\chi$, while neutral and single ionized iron lines may
provide the same abundance within the quoted errors.

Generally, the adopted temperature scale well satisfies the excitation
equilibrium and only a few field stars require re-adjusted temperatures. 
To better constrain the gravity values, we imposed the condition of [Fe/H]
\footnote{We adopt the usual spectroscopic notation: [A]=log$(A)_{star}$-log$(A)_{\odot}$ for any 
abundance quantity A; log(A) is the abundance by number of the element A in the standard scale 
where log(H)=12.}~I=[Fe/H]~II. Photometric and spectroscopic
gravities for the cluster stars are consistent, while for the
field stars we needed to re-adjust the gravities within $\pm$0.3 dex,
probably due to incorrect assumptions for their mass, reddening and/or distance modulus.

In order to estimate the micro-turbulent velocity $v_t$ we adopted
as {\sl initial} value a velocity of $v_t$=1.5 km $s^{-1}$ and we
adjusted this parameter in each star in order to minimize the
trend between [Fe/H]~I abundance and the expected line strength,
defined as $\lg{gf}-\theta\chi$ (where $\theta$ is
5040/$T_{eff}$), according to the prescriptions by \citet{magain}
and imposing in this way that strong and weak lines give
the same abundance.\\
The final atmospheric parameters (and the derived
[Fe/H] abundance ratios) are listed in Tab. \ref{param}.

Uncertainties in the derived atmospheric parameters have been 
computed by taking into account the main sources of errors. 
For $T_{eff}$, we considered uncertainties in the photometric 
(V-K) colors and reddening, finding uncertainties ranging from 
$\sim$70 to $\sim$120 K; in the following we assume a typical error 
of 100 K. 
The uncertainties in the gravities have been computed by considering 
the corresponding error in $T_{eff}$ (being log~g fixed by the choice of $T_{eff}$) 
and in the adopted reddening and mass. In particular, the error in the adopted mass 
is small for the cluster stars 
\citep[for which the age is well constrained, see e.g.][]{broc03}, while 
for the field stars we assume an error of the order of $\sim$30\%.
Typical errors in gravities are of the order of 0.2.
The errors in $v_t$ have been estimated by varying this parameter until the 
$\sigma_{slope}$ value for the slope in the line strength--A(Fe) plane is 
reached. Because $v_t$ is estimated spectroscopically, the associated errors 
depend on the SNR of the spectra and the number of adopted lines: 
we find that the errors in $v_t$ ranging from $\sim$0.15 km/s for the cluster stars 
to $\sim$0.3 km/s for the faintest field stars.

\begin{table}
\begin{center}
\caption{Atmospheric parameters and iron content for all
the target stars.} 
\begin{tabular}{ccccc}
\hline
ID-Star   & $T_{eff}$ & log~g &  $v_t$ &    [Fe/H]   \\
    & (K) &    & (km$s^{-1})$ & (dex)\\
\hline
        &         &    CLUSTER     &        &       \\
\hline
  2131  &   4080  &   1.05  & 2.0 &  --0.47	\\
  2981  &   3870  &   0.90  & 1.9 &  --0.45	\\
  4017  &   4490  &   1.70  & 1.8 &  --0.47	 \\
  4425  &   4530  &   1.45  & 1.8 &  --0.43	 \\
  4462  &   5320  &   1.90  & 1.7 &  --0.39	 \\
  5231  &   4100  &   0.90  & 2.1 &  --0.48	   \\
  5415  &   5540  &   2.05  & 1.5 &  --0.42	\\
  5706  &   4460  &   1.80  & 1.8 &  --0.38	\\
  5789  &   5110  &   1.90  & 1.5 &  --0.43    \\
  7392  &   5510  &   1.60  & 1.7 &  --0.38	\\
  7402  &   4900  &   2.10  & 1.5 &  --0.46	\\
  7415  &   5200  &   2.05  & 1.7 &  --0.49	 \\
  7862  &   4570  &   1.90  & 1.7 &  --0.46	\\
 10366  &   5760  &   2.20  & 1.7 &  --0.38	 \\
\hline
        &         &    FIELD     &        &       \\
\hline
   652  &   4530  &   1.90  & 1.4    & --0.71  \\
  1491  &   4760  &   2.00  & 1.5    & --0.44 \\
  1605  &   4360  &   1.50  & 1.5    & --0.85	  \\
  1995  &   4580  &   2.00  & 1.5    & --1.15 \\
  2305  &   4470  &   1.75  & 1.5    & --0.60 \\
  4209  &   4180  &   1.30  & 1.5    & --0.63	 \\
  7111  &   4550  &   1.90  & 1.4    & --0.59 \\
  9256  &   4870  &   2.30  & 1.6    & --0.33  \\
  9649  &   4660  &   2.05  & 1.4    & --0.32 \\
 10144  &   4390  &   1.80  & 1.4    & --0.75 \\
 10222  &   4420  &   1.75  & 1.3    & --0.52	 \\
\hline
\hline
\label{param}
\end{tabular}
\end{center}
\end{table}

\section{Chemical analysis}

For each star a plane-parallel, one-dimensional,
LTE model atmosphere has been computed by using the ATLAS~9 code
\citep{kur93a} in its Linux version \citep{sbordone04}
and adopting the atmospheric parameters described in Tab.~\ref{param}.
We used the {\sl new} Opacity Distribution Functions by \citet{castelli03},
with a solar-scaled chemical mixture 
\citep[according with the previous chemical analysis of NGC~1866 by][]{hill00}, 
micro-turbulent velocity
of 1 km $s^{-1}$, a mixing-length parameter of 1.25 and no
approximate overshooting.

\begin{figure}
\includegraphics[width=85mm]{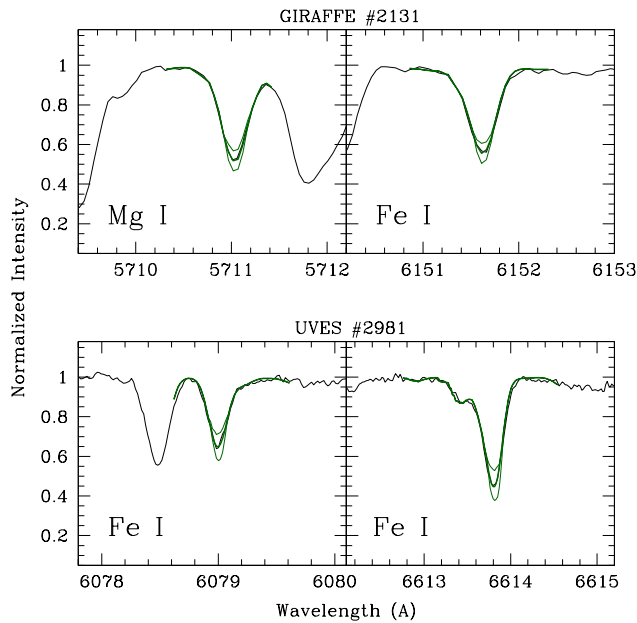}
\caption{Portions of spectrum for the two observed stars \#2131 (upper panels, GIRAFFE)
and \#2891 (lower panels, UVES) with overplotted the best fit (green, thick lines). 
Synthetic spectra with abundances of $\pm$0.1 dex with respect to the best fit 
spectra are also plotted as green, thin lines.}
\label{fit}
\end{figure}

For the chemical analysis of our sample we resort to the line
profile fitting technique, comparing the observed line profile
with suitable synthetic ones. The adopted code \citep[described in
detail in][]{caffau05} performs a $\chi^2$ minimization of the
deviation between synthetic profiles and the observed spectrum.
The best fitting spectrum is obtained by linear interpolation
between three synthetic spectra which differ only in the abundance
of a given element; the minimum $\chi^2$ is computed numerically
by using the MINUIT package \citep{james}. All the synthetic
spectra were computed with the SYNTHE code \citep{kur93b}. Fig.~\ref{fit} 
shows examples of final best-fit for used spectral
features in the GIRAFFE spectrum of the star \#2131 (upper panel)
and in the UVES spectrum of the star \#2981 (lower panel); 
synthetic spectra with abundances of $\pm$0.1 dex with 
respect to the best fit abundance are also plotted for sake of 
comparison.

We select a set of spectral lines (predicted to be un-blended
by the inspection of preliminary synthetic spectra computed with
the photometric atmospheric parameters) and adopting
accurate laboratory or theoretical oscillator strengths whenever
possible. In the computation of synthetic spectra we employ
the line-list of R. L. Kurucz database
\footnote{http://kurucz.harvard.edu/linelists/gf100/}, updating
the oscillator strengths where available. Hyperfine
splitting has been included for Mn~I, Cu~I, Ba~II, La~II and
Eu~II  lines. Briefly, we summary in the following the updated atomic data:\\
{\bf O I -- } for the forbidden [O] I transition at 6300.31
$\mathring{A}$ we use the \citet{storey} oscillator
strength, while for the blended Ni~I line at 6300.34 $\mathring{A}$
we adopt the \citet{j03} laboratory log~gf; \\
{\bf Mg I -- } we use the \citet{gr03} log~{\sl gf} 
for the Mg~I transitions at 5711.09, 6318.71 and 6319.24 $\mathring{A}$;\\
{\bf Mn I -- } hyperfine splitting from
R. L. Kurucz website
\footnote{http://kurucz.harvard.edu/linelists/gfhyper100/}
 are employed;\\
{\bf Cu II --} for the 5782.0 $\mathring{A}$ line the hyperfine
levels are from \citet{cunha}, adopting a solar isotopic mixture;\\
{\bf Ba II -- } we use the hyperfine components by \citet{prochaska}
for the Ba~II lines at 5853.7, 6141.6 and 6496.9 $\mathring{A}$;\\
{\bf Rare earths --} the transition probability of the 6043.4 Ce~II line
is from DREAM Database \footnote{http://w3.umh.ac.be/~astro/dream.shtml}
and of the 5740.8 Nd~II line by \citet{den};\\
{\bf La II and Eu II -- } hyperfine splitting is included,
by adopting the recent atomic data
by \citet{law1} and \citet{law2} for Eu~II and La~II respectively. 
We perform the calculation of their hyperfine structure with the
LINESTRUC code, described by \citet{wal}.

The Na lines are affected by NLTE effects and such
corrections are a function of line strength, metallicity,
temperature and gravity. We correct our Na abundances for
departures from LTE, interpolating the grid by \citet{gr99}.

All the abundances are referred to the solar values listed in the
recent compilation by \citet{lodders}, adopting only for O and Eu
the new solar abundances by \citet{caffau08} and \citet{mucc08},
respectively, and for Mg, Al and Cu the values derived from our 
solar analysis.
For sake of homogeneity, we perform an analysis of the solar spectrum
by using the same procedure adopted here. We study the Kurucz flux
spectrum \footnote{See http://kurucz.harvard.edu/sun.html} and employ
the ATLAS~9 solar model atmosphere computed by F. Castelli
\footnote{http://www.user.oats.inaf.it/castelli/sun/ap00t5777g44377k1asp.dat}.
Generally, we find that our solar analysis nicely agrees with the
solar values by \citet{lodders} within the uncertainties. We note that
only for few elements there are relevant differences with respect to
the values by \citet{lodders}.
Our solar Mg abundance is of 7.43, while \citet{lodders} recommended value
is of of 7.54; such a discrepancy on the line selection can be attributed to the
adopted log~gf, as discussed by \citet{gr03}.
Al abundance is of 6.21 from the doublet at 6696--98 $\mathring{A}$
(0.26 dex lower than the value listed by \citet{lodders}), probably due
to NLTE effects that affect these lines and/or imprecise log~gf values
\footnote{It is worth noting that such a discrepancy in solar Al abundance
has been revealed by other authors, see e.g. \citet{reddy} and \citet{gr03}.}.
Finally, our Cu solar abundance is 0.2 dex lower than the reference value.
Such a difference has been already noted by \citet{cunha} and ascribed
to the differing log~gf values and model atmospheres.

\section{Error budget}

In the case of observed spectra, where adjacent pixels are not
completely independent of each other, the error associated to the
$\chi^2$ minimization cannot be derived by the $\chi^2$ theorems
\citep[see][]{cayrel99, caffau05}. In order to estimate the
uncertainties related to the fitting procedure we resort to Monte
Carlo simulations. We choose to study some cluster stars, which we
consider as representative of the different S/N and atmospheric
parameters sampled by our targets: the stars \#2131 and \#10366,
located in the red giant region and in the blue side of the Blue
Loop of NGC~1866, respectively, and the field RGB star \#652.
We injected Poisson noise into the best-fit synthetic spectrum of
some iron lines, according to the standard deviation used in the
fitting and we performed the fit with the same procedure described
above. For each line we performed a total of 10000 Monte Carlo
events. From the resulting abundance distributions we may estimate
a 1$\sigma$ level for normal distributions. The two cluster stars
exhibit similar Monte Carlo distributions. We claim that the
abundances derived by our fitting procedure are constrained within
$\pm$0.09 dex. We repeated the same procedure for \#652 (the star
with the lowest S/N of the sample, S/N=~40), estimating that the
68\% of the events is comprised within 0.15 dex.

We computed for the stars \#2131 and \#10366 the sensitivity
of each abundance ratio to variation of the atmospheric parameters.
We assume typical errors for each parameter according to Section 3.
Tab. \ref{error} lists the variations of the abundance ratios
by varying each time one only parameter and their sum in quadrature
can be considered a conservative estimate of the systematic error
associated to a given abundance ratio.

\begin{table*}
\begin{minipage}{120mm}
\caption{Variations in the abundances of two stars \#2131 and \#10366
due to the uncertainties in the atmospheric parameters. The adopted 
parameters variations are also reported.}
\begin{tabular}{ccccccc}
\hline
&   &  \#2131   &  & &  \#10366 & \\
\hline
 Ratio   &  $T_{eff}$&
log~g  &  $v_t$ & $T_{eff}$&
log~g  &  $v_t$  \\
 & (100 K) & (0.2) & (0.3 km/s) & (100 K) & (0.2) & (0.3 km/s) \\
\hline
$[Fe/H] $  & --0.06  &  0.02 & -0.03  &--0.04  &  --0.01  &--0.05    \\
$[Na/Fe]$  & --0.05  &  0.03 &--0.08  &--0.03  &  --0.03  &--0.10   \\
$[O/Fe] $  &   0.05  &--0.04 &--0.05  &  0.04  &  --0.05  &--0.07   \\
$[Mg/Fe]$  & --0.04  &  0.04 &  0.04  &  0.04  &    0.01  & 0.02  \\
$[Si/Fe]$  & --0.06  &  0.03 &  0.02  &--0.05  &    0.04  & 0.03  \\
$[Ca/Fe]$  &   0.02  &--0.06 &  0.03  &  0.05  &  --0.06  & 0.04     \\
$[Ti/Fe]$  &   0.12  &  0.03 &  0.10  &  0.11  &  --0.01 & 0.12    \\
$[Mn/Fe]$  & --0.15  &  0.04 &  0.08  &--0.08  &    0.06 & 0.12    \\
$[Ni/Fe]$  &  -0.03  &--0.02 &  0.02  &--0.02  &    0.02 & 0.03    \\
$[Cu/Fe]$  &  -0.07  &  0.06 &  0.08  &  0.05  &    0.09 & 0.05   \\
$[Y/Fe] $  & --0.04  &  0.07 &  0.04  &  0.03  &    0.05  & 0.03  \\
$[Zr/Fe]$  &   0.14  &  0.04 &  0.03  &  0.12  &  --0.04  & --0.02 \\
$[Ba/Fe]$  &   0.04  &  0.06 &  0.12  &  0.05  &    0.08 &  0.10  \\
$[La/Fe]$  &   0.02  &  0.03 &  0.04  &  0.03  &  --0.02  &  0.01  \\
$[Ce/Fe]$  &   0.02  &  0.03 &--0.02  &--0.04  &   -0.01 &  0.01 \\
$[Nd/Fe]$  &   0.03  &  0.08 &  0.03  &  0.01  &    0.06 &  0.04  \\
\hline
\label{error}
\end{tabular}
\end{minipage}
\end{table*}

\section{Results}

Tab. \ref{cl} and \ref{fld} list the derived abundance ratios for all
the samples of stars (cluster and field respectively) and
Tab. \ref{aver} the average values (with the corresponding
dispersion by the mean) obtained for NGC~1866.
Two of the targets (namely \#1146 and \#5834) are affected by strong TiO bands, 
thus have not been analyzed due to the severe molecular absorption conditions.
It is worth noting that the dispersion by
the mean for each abundance ratio in NGC~1866 is consistent within the uncertainties
arising from the fitting procedure and the atmospheric parameters,
pointing toward a general homogeneity for all the studied elements 
based on more than a single star (see Section 6.5).

\begin{figure}
\includegraphics[width=90mm]{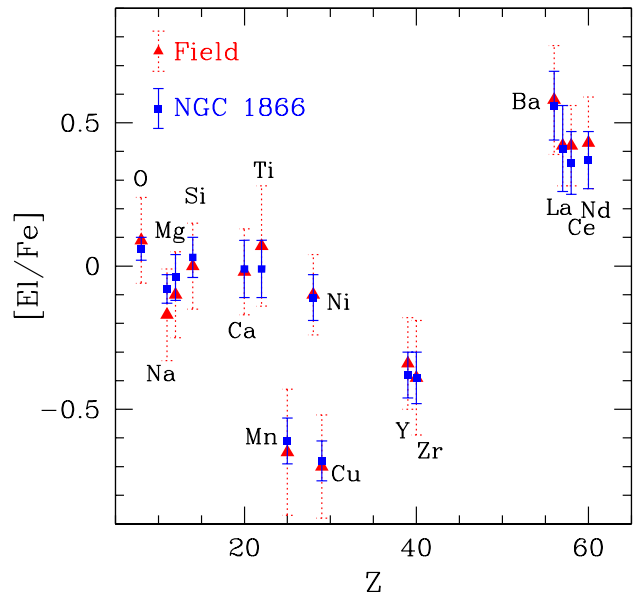}
\caption{Comparison between the mean spectroscopic values of stars
belonging to NGC~1866 (blue squares) and the surrounding field (red
triangles). Errorbars indicate the dispersion by the mean.}
\label{average}
\end{figure}

In Fig.~\ref{average} a full picture of the chemical abundances
inferred from our sample is shown: blue squares are the average
values for NGC~1866 and red triangles for the LMC field stars. In
Fig.~\ref{ona}--\ref{bala} we summarize the derived abundances of our sample for
some interesting elements (filled grey points for the field stars
and grey large square for the average value of the stars of NGC~1866),
comparing these results with other databases based on
high-resolution spectroscopy for the Galactic stars 
\citep[empty grey points, by][]{edv, burris, ful, reddy,gr03,reddy06}, the LMC
field stars \citep[blue points by][]{smith,pomp} and the LMC
globular clusters \citep[blue squares by][]{jo06,m08,m10}. 

\begin{table*}
\begin{minipage}{150mm}
\caption{Abundances ratios for the target stars of NGC~1866. The numbers in brackets indicate 
the number of used lines.}
\label{anymode}
\begin{tabular}{ccccccccc}
\hline
ID-Star   & [Na/Fe]& [O/Fe]
 &  [Mg/Fe]  & [Si/Fe]&
[Ca/Fe]   & [Ti/Fe]  &
[Ni/Fe]   & [Mn/Fe] \\
 & (dex) & (dex) &   (dex) & (dex) & (dex) & (dex) &   (dex) & (dex) \\
\hline
  2131  & --0.09~(4) &   0.11~(1)   &	  0.03~(3)   &     0.02~(4)  &  --0.13~(8)  &  --0.08~(8) &   --0.13~(10) &  --0.58~(3)  \\
  2981  & --0.12~(4) &   0.10~(1)   &   --0.04~(3)   &     0.09~(4)  &  --0.12~(10) &  --0.13~(11)&   --0.05~(11) &  --0.69~(3)  \\
  4017  & --0.07~(4) &	 0.01~(1)   &   --0.12~(3)   &	   0.03~(6)  &  --0.02~(8)  &  --0.05~(10 &   --0.16~(8)  &  --0.55~(3)  \\
  4425  & --0.11~(4) &	 0.13~(1)   &   --0.09~(3)   &	   0.09~(5)  &    0.05~(6)  &    0.14~(6) &   --0.13~(8)  &  --0.56~(3)  \\
  4462  & --0.03~(4) &	 0.09~(1)   &	  0.02~(3)   &   --0.04~(4)  &  --0.01~(8)  &  --0.02~(6) &     0.04~(12) &  --0.55~(3)  \\
  5231  & --0.13~(4) &	 0.00~(1)   &   --0.01~(3)   &   --0.07~(5)  &  --0.16~(9)  &  --0.04~(8) &   --0.17~(10) &  --0.61~(3) \\
  5415  & --0.11~(4) &	 0.03~(1)   &   --0.08~(3)   &	   0.20~(5)  &    0.14~(8)  &    0.25~(8) &   --0.20~(8)  &  --0.63~(3)  \\
  5706  & --0.19~(4) &	 0.11~(1)   &   --0.03~(3)   &     0.08~(5)  &  --0.17~(7)  &  --0.15~(8) &   --0.12~(7)  &  --0.66~(3)  \\
  5789  & --0.02~(4) &	 0.07~(1)   &   --0.07~(3)   &   --0.06~(4)  &    0.10~(8)  &  --0.03~(7) &   --0.03~(6)  &  --0.81~(3)  \\
  7392  & --0.12~(4) &	 0.04~(1)   &	  0.10~(3)   &   --0.02~(5)  &    0.11~(6)  &  --0.03~(8) &   --0.12~(8)  &  --0.62~(3)  \\
  7402  & --0.10~(4) &	 0.09~(1)   &   --0.17~(3)   &	   0.03~(5)  &    0.04~(6)  &    0.05~(9) &   --0.04~(8)  &  --0.60~(3)  \\
  7415  & --0.04~(4) &	 0.06~(1)   &	  0.02~(3)   &     0.06~(4)  &  --0.12~(7)  &    0.00~(5) &   --0.13~(11) &  --0.51~(3)  \\
  7862  & --0.11~(4) &	 0.10~(1)   &   --0.16~(3)   &	   0.08~(4)  &  --0.01~(8)  &  --0.04~(6) &     0.00~(10) &  --0.64~(3)  \\
 10366  & --0.02~(4) &	 0.02~(1)   &   --0.09~(3)   &	   0.07~(5)  &    0.00~(8)  &  --0.04~(8) &   --0.23~(9)  &  --0.48~(3)  \\
\hline
ID-Star   &  [Cu/Fe] &   [Y/Fe]&
[Zr/Fe]  &  [Ba/Fe]&
[La/Fe] & [Ce/Fe]&
[Nd/Fe]  & [Fe/H] \\
 & (dex) & (dex) &   (dex) & (dex) & (dex) & (dex) &   (dex) & (dex) \\
\hline
 2131  &    --0.76~(1)	&  --0.22~(2)	&    --0.52~(3)     &   0.52~(2)      &	0.37~(1)   &  0.25~(1)  &    0.51~(3)   &   --0.47~(42)   \\
 2981  &     ---   	&  --0.45~(5)	&    --0.51~(4)     &   0.54~(3)      &	0.44~(1)   &  0.41~(3)  &    0.52~(8)   &   --0.45~(89)   \\
 4017  &    --0.67~(1)	&  --0.39~(1)	&    --0.21~(3)     &   0.55~(2)      &	0.60~(1)   &  0.20~(1)  &    0.37~(3)   &   --0.47~(40)   \\
 4425  &    --0.69~(1)	&  --0.33~(2)	&    --0.41~(3)     &   0.63~(2)      &	0.33~(1)   &  0.29~(1)  &    0.24~(3)   &   --0.43~(38)   \\
 4462  &    --0.70~(1)	&  --0.33~(2)	&      --- 	    &   --- 	      &	0.40~(1)   &  0.25~(1)  &    0.38~(3)   &   --0.39~(44)   \\
 5231  &    --0.70~(1)	&  --0.53~(1)	&    --0.49~(3)     &   0.51~(2)      &	0.36~(1)   &  0.17~(1)  &    0.47~(2)   &   --0.48~(40)   \\
 5415  &    --0.69~(1)	&  --0.36~(2)	&    --0.38~(2)     &   ---           &	0.18~(1)   &  0.41~(1)  &    0.23~(3)   &   --0.42~(37)   \\
 5706  &    --0.75~(1)	&  --0.49~(2)	&    --0.46~(3)     &   0.48~(2)      &	0.35~(1)   &  0.28~(1)  &    0.24~(2)   &   --0.38~(39)   \\
 5789  &    --0.57~(1)	&   --- 	&    --0.33~(3)     &   0.64~(2)      &	0.20~(1)   &  0.44~(1)  &   ---         &   --0.43~(39)   \\
 7392  &    --0.60~(1)	&  --0.44~(2)	&      ---          &   0.61~(2)      &	0.18~(1)   &  0.19~(1)  &    0.38~(2)   &   --0.38~(42)   \\
 7402  &    --0.58~(1)	&  --0.43~(2)	&    --0.40~(3)     &   0.58~(2)      &	0.39~(1)   &  0.20~(1)  &    0.45~(3)   &   --0.46~(40)   \\
 7415  &    --0.82~(1)	&  --0.38~(2)	&    --0.44~(3)     &   0.55~(2)      &	0.60~(1)   &  0.27~(1)  &    0.32~(3)   &   --0.49~(42)   \\
 7862  &    --0.71~(1)	&  --0.43~(2)	&    --0.42~(3)     &   0.46~(2)      &	0.42~(1)   &  0.17~(1)  &    0.34~(3)   &   --0.46~(37)   \\
10366  &     ---  	&  --0.42~(1)	&      --- 	    &   0.62~(2)      &	0.67~(1)   &  0.51~(1)  &    0.36~(3)   &   --0.38~(40)   \\
\hline
ID-Star   &  [Al/Fe] &   [Mo/Fe]&
[Ru/Fe]  &  [Hf/Fe]&
[W/Fe] & [Pr/Fe]&
[Eu/Fe]  &  [Er/Fe] \\
 & (dex) & (dex) &   (dex) & (dex) & (dex) & (dex) &   (dex) &  (dex)\\
\hline
  2981  &  --0.30~(2)  &  --0.03~(2)  &   --0.05~(1)  & 0.17~(2)  & 0.02~(1)  & 0.51~(5)  &  0.57~(1)  & 0.30~(2) \\
\hline
\label{cl}
\end{tabular}
\end{minipage}
\end{table*}

\begin{table*}
\begin{minipage}{150mm}
\begin{center}
\caption{Abundance ratios of the LMC field target stars. The numbers in brackets indicate 
the number of used lines.}
\label{anymode}
\begin{tabular}{ccccccccc}
\hline
ID-Star   & [Na/Fe]& [O/Fe]
 &  [Mg/Fe]  & [Si/Fe]&
[Ca/Fe]   & [Ti/Fe]  &
[Ni/Fe]   & [Mn/Fe] \\
 & (dex) & (dex) &   (dex) & (dex) & (dex) & (dex) &   (dex) & (dex) \\
\hline
652    &   --0.12~(4) &     0.12~(1) &	  0.02~(3)     &       0.11~(5)	&  --0.05~(7)	 &  --0.08~(7)	 &  --0.14~(8)	  &  --0.65~(3)    \\
1491   &   --0.31~(4) &      ---     &	--0.14~(3)     &       0.04~(8)	&  --0.03~(8)	 &  --0.09~(7)	 &  --0.12~(7)	  &  --0.70~(3)    \\
1605   &   --0.04~(4) &     0.12~(1) &	--0.21~(3)     &     --0.05~(6)	&  --0.07~(6)	 &    0.14~(7)	 &    0.10~(8)	  &  --0.80~(3)    \\
1995   &   --0.25~(4) &     0.17~(1) &	  0.07~(3)     &       0.08~(3)	&    0.12~(4)	 &    0.20~(4)	 &  --0.20~(5)	  &  --0.75~(3)    \\
2305   &   --0.12~(4) &     0.07~(1) &	--0.04~(3)     &     --0.04~(5)	&  --0.02~(8)	 &    0.12~(8)	 &  --0.09~(8)	  &  --0.57~(3)    \\
4209   &   --0.26~(4) &     0.09~(1) &	  0.03~(3)     &     --0.04~(5)	&  --0.04~(6)	 &    0.20~(9)	 &  --0.06~(6)	  &  --0.63~(3)    \\
7111   &   --0.22~(4) &      ---     &	--0.16~(3)     &     --0.11~(8)	&    0.04~(7)	 &    0.10~(6)	 &  ---    	  &  --0.64~(3)    \\
9256   &   --0.19~(4) &    ---       &	--0.17~(3)     &       0.03~(7)	&    0.00~(7)	 &    0.01~(7)	 &  --0.13~(7)	  &  --0.52~(3)    \\
9649   &     ---     &    --0.03~(1) &	--0.18~(3)     &     --0.04~(6)	&  --0.07~(8)	 &    0.05~(6)	 &  --0.11~(7)	  &  --0.54~(3)    \\
10144  &   --0.25~(4) &     0.20~(1) &	  0.01~(3)     &     --0.10~(7)	&    0.02~(8)	 &    0.24~(8)	 &  --0.01~(7)	  &  --0.76~(3)    \\
10222  &   --0.22~(4) &      ---     &	--0.10~(3)     &     --0.08~(7)	&  --0.01~(8)	 &    0.00~(7)	 &  --0.02~(8)	  &  --0.56~(3)    \\
\hline
ID-Star   &  [Cu/Fe] &   [Y/Fe]&
[Zr/Fe]  &  [Ba/Fe]&
[La/Fe] & [Ce/Fe]&
[Nd/Fe]  &  [Fe/H] \\
 & (dex) & (dex) &   (dex) & (dex) & (dex) & (dex) &   (dex) & (dex) \\
\hline
 652     &    --0.69~(1)   &   ---	    &    --0.30~(3)      &     0.60~(2)	&   0.54~(1)   &     0.54~(1)   &   0.26~(3)	 &  --0.71~(40)  \\
1491     &    --0.76~(1)   &  --0.51~(2)    &    --0.55~(2)      &     0.64~(2)	&   0.58~(1)   &    ---         &   0.12~(3)	 &  --0.44~(42)  \\
1605     &    --0.92~(1)   &  ---	    &    --0.27~(3)      &     0.73~(2)	&   0.29~(1)   &    ---         &   0.49~(2)	 &  --0.85~(36)  \\
1995     &    --1.11~(1)   &  --0.16~(2)    &    --0.21~(2)      &     0.28~(2)	&   0.18~(1)   &     0.13~(1)    &   ---  	 &  --1.15~(32)  \\
2305     &    --0.65~(1)   &  --0.34~(1)    &    --0.26~(3)      &     0.62~(2)	&   0.23~(1)   &    ---         &   0.50~(3)	 &  --0.60~(41)  \\
4209     &    --0.77~(1)   &  --0.34~(1)    &    --0.32~(3)      &     0.40~(2)	&   0.53~(1)   &     0.56~(1)   &   0.38~(2)	 &  --0.63~(40)  \\
7111     &    --0.59~(1)   &	---	    &    --0.35~(3)      &     0.58~(2)	&   0.51~(1)   &     0.37~(1)   &   0.55~(3)	 &  --0.59~(35)  \\
9256     &    --0.60~(1)   &  --0.34~(2)    &    --0.50~(3)      &     0.54~(2)	&   0.32~(1)   &     0.53~(1)   &   0.54~(3)	 &  --0.33~(38)  \\
9649     &    --0.69~(1)   &  ---	    &    --0.52~(3)      &     0.51~(2)	&   0.54~(1)   &     0.35~(1)   &   0.45~(2)	 &  --0.32~(40)  \\
10144    &    --0.76~(1)   &  --0.14~(2)    &    --0.09~(2)      &     0.60~(2)	&   0.58~(1)   &     0.48~(1)   &   --- 	 &  --0.75~(34)  \\
10222    &    --0.49~(1)   &  ---	    &    --0.45~(2)      &     0.73~(2)	&   0.45~(1)   &     0.53~(1)   &   0.27~(3)	 &  --0.52~(35)  \\
\hline
\label{fld}
\end{tabular}
\end{center}
\end{minipage}
\end{table*}

\subsection{The iron abundance}
We derived an average iron content for NGC~1866 of
[Fe/H]=~--0.43$\pm$0.01 dex ($\sigma$=~0.04 dex). This abundance
agrees with the previous one by \citet{hill00} from the analysis
of 3 giants, with [Fe/H]=~--0.50$\pm$0.03 dex ($\sigma$=~0.06 dex).
The small offset between the two iron determinations can be
ascribed to the different model atmospheres adopted and reference
solar values (the \citet{lodders} solar iron abundance is 0.04 dex lower
than the \citet{gs98} value). The iron abundance of NGC~1866 
agrees with the metallicity of the intermediate-age LMC clusters 
by \citet{m08}.
On the other side, recently \citet{colucci} derived a higher 
([Fe/H]=~+0.04$\pm$0.04 dex) iron abundance for the cluster, by using 
high-resolution integrated spectra. At present, we have not all the details 
of their analysis and we cannot identify the origin of the discrepancy.
[Fe/H] of field stars ranges
from --1.15 to --0.32 dex, in agreement with the metallicity distribution
for the LMC stars derived by \citet{cole} and \citet{pomp}.

\subsection{O and Na}
\begin{figure}
\includegraphics[width=85mm]{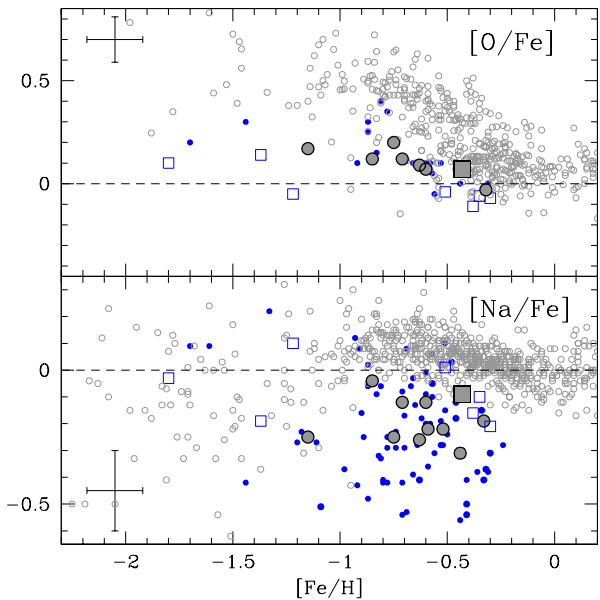}
\caption{Behaviour of [O/Fe] (upper panel) and [Na/Fe] (lower panel) as a function
of [Fe/H] for the observed stars: the grey square is the average value for the stars
of NGC~1866, large grey points the
individual LMC field stars, blue squares the intermediate-age LMC clusters by
\citet{m08} and the old LMC clusters by \citet{jo06} and \citet{m10},
the small grey points Galactic stars by \citet{edv, ful, burris, reddy, gr03, reddy06}
and the small blue points the LMC field giants by \citet{pomp} and \citet{smith}.
Errorbars indicate the typical uncertainties arising from
the atmospheric parameters and the error in the fitting procedure.}
\label{ona}
\end{figure}

\begin{figure}
\includegraphics[width=85mm]{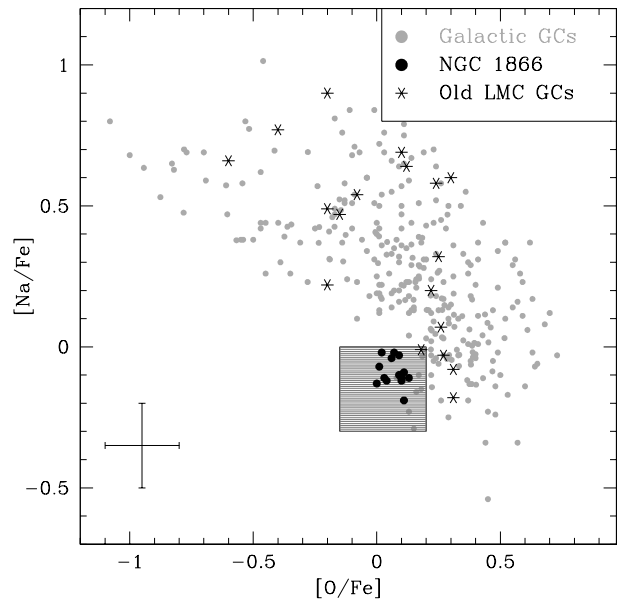}
\caption{Behaviour of [Na/Fe] ratio as a function of [O/Fe] for the individual stars
of NGC~1866 (black points). In comparison the individual stars observed in Galactic GCs (grey
points) and in the old LMC GCs \citep[black asterisks, by][]{m10} have been plotted. 
Light grey area indicates the mean locus defined by the stars measured by \citet{m08} 
in 4 intermediate-age LMC clusters}.
\label{anticor}
\end{figure}

Stars of NGC~1866, as well as the field stars of our sample, show
[O/Fe] and [Na/Fe] abundance ratios generally lower than the
Galactic stars (see Fig.~\ref{ona}). The average [O/Fe] ratio for
NGC~1866 is of +0.07 dex ($\sigma$=~0.04 dex), while the [Na/Fe] derived
is of --0.09 dex ($\sigma$=~0.05 dex).  
We note quite different [Na/Fe] abundances in our stars with respect to 
the sample of LMC field stars by \citet{pomp}: basically, their [Na/Fe] abundances 
range from --0.6 up to +0.2 dex, while our measures share a typical value of $\sim$--0.2 dex.
Note that their Na abundances do not include corrections for 
departures from LTE conditions, at variance with our analysis. In fact, NLTE 
corrections depend simultaneously on temperature, metallicity, gravity and line strength, 
and the choice to neglect these effects can enlarge the star-to-star Na differences.
In contrast to the
observational evidences in the Galactic GCs studied so far (where relevant
star-to-star variations in O and Na abundance have been revealed),
the O/Na content of NGC~1866 appears to be homogeneous and
the observed scatters are consistent within the quoted
uncertainties. Fig.~\ref{anticor} reports in the [O/Fe]-[Na/Fe]
plane the individual stars of NGC~1866 (black points), in comparison with the
individual stars observed in several Galactic GCs (grey points) and in the 
old LMC GCs by \citet{m09}. The grey region indicates the mean locus of the 
giant stars in intermediate-age LMC clusters by \citet{m08}.

\subsection{$\alpha$-elements}
\begin{figure}
\includegraphics[width=85mm]{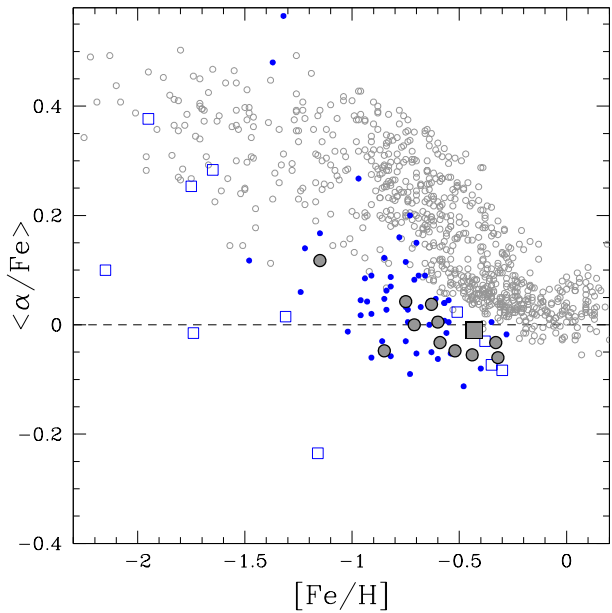}
\caption{Behaviour of the [$\alpha$/Fe] ratio (defined as [Mg+Si+Ca+Ti/Fe]/4)
as a function of [Fe/H]. Same symbols of Fig.~\ref{ona}.}
\label{alfa1}
\end{figure}

For the other $\alpha$-elements (namely, Mg, Si, Ca and Ti)
NGC~1866 displays solar-scaled patterns, in a similar fashion of
the field giants. Fig.~\ref{alfa1} shows $<$$\alpha$/Fe$>$ (defined as mean
of [Mg/Fe], [Si/Fe], [Ca/Fe] and [Ti/Fe]) as a function of [Fe/H]: a mild
trend with the metallicity seems to be observed.
$<$$\alpha$/Fe$>$ ratios in both NGC~1866 and the LMC field stars appear to be lower 
than those observed in the Galactic stars at the same metallicity level; the same result has 
been pointed out by \citet{pomp}. At lower metallicities 
([Fe/H]$<$--1 dex) the comparison between the LMC and the Galaxy is quite complex.
In fact, the old LMC clusters by \citet{m10} exhibit a quite good agreement 
with the Galactic Halo stars, while the clusters analyzed by \citet{jo06} 
show systematically lower [Ti/Fe] and [Ca/Fe] ratios, but similar [Si/Fe] ratios. 
Note that the sample of LMC field stars discussed here does not include stars 
with [Fe/H]$<$-1.5 dex and does not allow to identify possible discrepancy between 
the [$\alpha$/Fe] ratio between the Halo stars and the metal-poor component of the LMC.

\subsection{Mn, Cu and Ni}
\begin{figure}
\includegraphics[width=85mm]{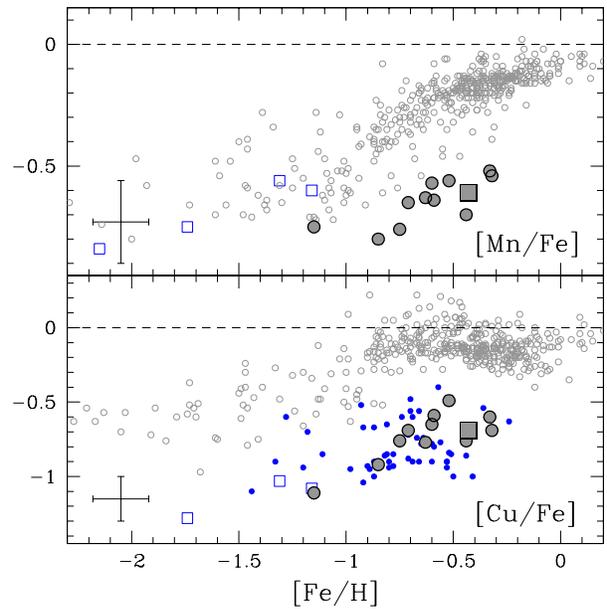}
\caption{Behaviour of the [Mn/Fe] (upper panel) and
[Cu/Fe] (lower panel)
as a function of [Fe/H]. Same symbols of Fig.~\ref{ona}.}
\label{mncu}
\end{figure}

Both [Mn/Fe] and [Cu/Fe] abundance ratios in our sample display 
significant underabundances with respect to the Galactic patterns (see
Fig.~\ref{mncu}). We found for NGC~1866 average values of
[Mn/Fe]=~--0.61 dex ($\sigma$=~0.08 dex) and [Cu/Fe]=~--0.69 dex
($\sigma$=~0.07 dex). Such a depletion has been detected also
in the LMC field stars that exhibit a clear trend of decreasing [Mn/Fe] and [Cu/Fe]
with the metallicity.
Ni abundances are [Ni/Fe]=~--0.10 ($\sigma$=~0.08 dex) and [Ni/Fe]=~--0.08 
($\sigma$=~0.08 dex) for cluster and field stars respectively.

\subsection{Neutron-capture elements}
\begin{figure}
\includegraphics[width=85mm]{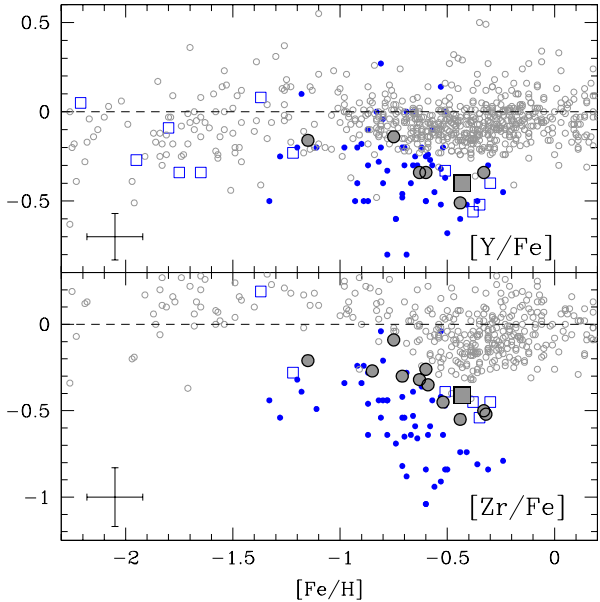}
\caption{Behaviour of the [Y/Fe] (upper panel) and
[Zr/Fe] (lower panel)
as a function of [Fe/H]. Same symbols of Fig.~\ref{ona}.}
\label{yzr}
\end{figure}

\begin{figure}
\includegraphics[width=85mm]{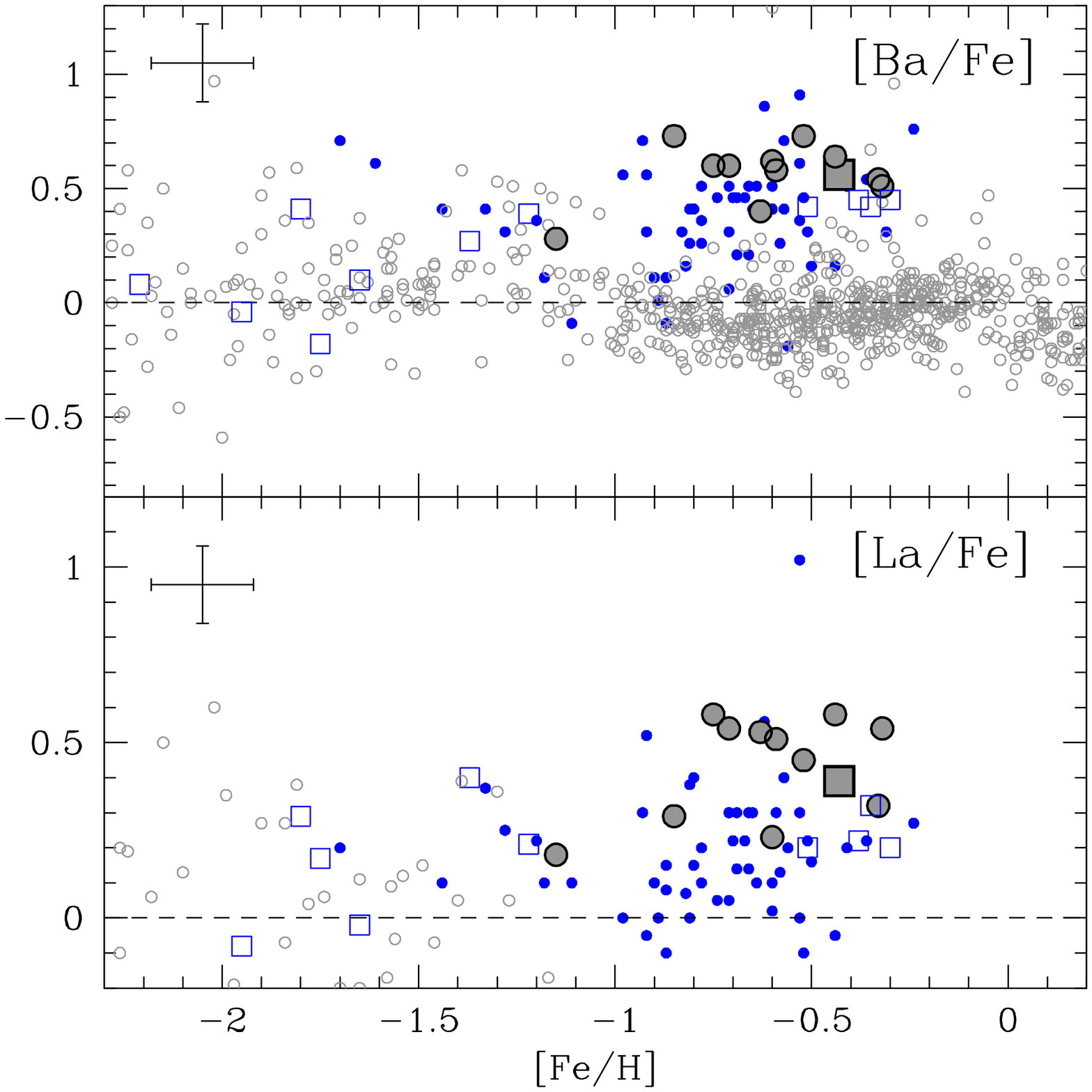}
\caption{Behaviour of the [Ba/Fe] (upper panel) and [La/Fe]
(lower panel)
as a function of [Fe/H]. Same symbols of Fig.~\ref{ona}.}
\label{bala}
\end{figure}

The elements belonging to the first peak of the s-elements, as Y
and Zr, turn out to be depleted with respect to the solar value
(Fig.~\ref{yzr}): we found for NGC~1866 average values of
[Y/Fe]=~--0.40 dex ($\sigma$=~0.08 dex) and [Zr/Fe]=~--0.41 dex
($\sigma$=0.09 dex), that well resemble the observed patterns in
the field stars. On the other hand, we 
detected enhanced abundance ratios for the second s-peak elements Ba, 
La, Ce and Nd (see
Fig.~\ref{bala}). 
We note a general offset between our abundances of 
[Zr/Fe] and [La/Fe] and the abundances by \citet{pomp}, while 
for [Y/Fe] and [Ba/Fe] the two samples well agree. The origin 
of the discrepancy is likely due to the use of different transitions between 
the two works. Each GIRAFFE setup covers only a rather small wavelength coverage 
and we have observed different GIRAFFE setups than \citet{pomp}. The use of 
different lines may bring some systematic offset in the retrieved abundances. This 
is usually averaged out by using many transitions, but residual differences may 
be present for those elements for which few transitions are available.

Abundances of other elements (namely Mo, Ru, Pr,
Eu, Er, Hf and W) have been measured only for the star \#2981 (see
Tab.~\ref{cl}), due to the large wavelength coverage of UVES. 
In particular, europium shows an enhanced value of [Eu/Fe]=~+0.49
dex.

\begin{table}
\begin{center}
\caption{Average abundance ratios for NGC~1866 and
corresponding standard deviation.}
\label{anymode}
\begin{tabular}{ccc}
Ratio   & Average&
$\sigma$\\
   &  (dex) & (dex) \\
\hline 
$[Fe/H] $  &  --0.43	&   0.04     \\ 
$[Na/Fe]$  & --0.09    &   0.05    \\ 
$[O/Fe] $  &   0.07   &   0.04    \\ 
$[Mg/Fe]$  &  --0.05   &   0.08   \\ 
$[Si/Fe]$  &  0.04    &   0.07   \\ 
$[Ca/Fe]$  & --0.02    &   0.10      \\ 
$[Ti/Fe]$  & --0.01    &   0.10    \\ 
$[Mn/Fe]$  & --0.61    &   0.08    \\  
$[Ni/Fe]$  & --0.10    &   0.08    \\ 
$[Cu/Fe]$  & --0.69    &   0.07   \\ 
$[Y/Fe] $  & --0.40    &   0.08   \\ 
$[Zr/Fe]$  & --0.41    &   0.09   \\ 
$[Ba/Fe]$  &  0.56    &   0.06  \\ 
$[La/Fe]$  &  0.39    &   0.15  \\ 
$[Ce/Fe]$  &  0.29    &   0.11 \\ 
$[Nd/Fe]$  &  0.37    &   0.10  \\ 
\hline
\label{aver}
\end{tabular}
\end{center}
\end{table}

\section{Discussion} \label{discu}
 
The Star Formation History (SFH) of irregular galaxies like the LMC 
is deeply different from the Milky Way; it is thought
to develop slowly, with several, short bursts of star formation, followed 
by long quiescent periods.
The theoretical interpretation of the chemical patterns in stars belonging to LMC
requires therefore some important caveats; in particular, we stress
the role that dynamical environmental processes (such as tidal
interaction and/or ram pressure stripping) may have on the
chemical evolution of a galaxy (see, e.g., \citet{bek} and
references therein). Indeed, \citet{besla} have suggested 
that the LMC entered the Galactic virial radius $\sim$3 Gyr ago, and 
tidal interactions with the Galaxy and the Small Magellanic Cloud likely triggered 
star formation that appears to have lasted $\sim$1 Gyr following that event.
In our analysis we do not account
for such effects.\\ 
As it is well known, main classes of chemical polluters are: 
\begin{itemize} 
\item{SuperNovae of type Ia (SN Ia), responsible for a large
production of iron and iron-peak elements;}  
\item{SuperNovae of type II (SN II), 
which synthesize oxygen, $\alpha$ elements, iron and iron-peak
elements, elements belonging to the weak component of the
s-process\footnote{These objects, in fact, efficiently synthesize
intermediate mass elements (ranging from copper to zirconium)
during their core He-burning and their C-shell burning.} and the
r-process elements;} 
\item{asymptotic giant branch (AGB) stars,
which pollute the Interstellar Medium (ISM) with carbon and
elements belonging to the main component of the
s-process\footnote{These elements are
commonly grouped in {\it ls} (light s) elements (Sr,Y,Zr) and {\it
hs} (heavy s) elements (Ba,La,Ce,Nd,Sm), representing the first
and the second peak of the s-process, respectively. Lead, which is
the termination-point of the s-process, constitutes the third
s-process peak.}.}
\end{itemize}
At the moment, the exact stellar site in which the r-process takes
place is still a matter of debate: this fact leads to strongly
different nucleosynthetic paths depending on the adopted physics
and theoretical assumptions \citep{qw,kra}. More robust
theoretical predictions are available for the s-process
\citep{ga98,busso,cri09}, which characterizes the thermally
pulsing phase of low mass AGB stars (TP-AGB phase).

In the following, we discuss three main aspects of our results: 
(i)~the internal abundance scatter of the stars in NGC~1866, in light 
of the self-enrichment scenario invoked to explain the internal
abundance spread of the old GCs; (ii)~possible chemical variations 
due to the different evolutive stages of the observed stars in this work; 
(iii)~the chemical abundances of NGC~1866 and its surrounding field 
in light of the chemical evolution of the LMC.


\subsection{NGC~1866 internal abundance scatter}

Before analyzing the spectroscopic patterns of single stars
belonging to the cluster, it is useful to compare abundances of
cluster stars with respect to stars lying in the surrounding
field. From Fig.~\ref{average}, in which we report mean values for
NGC~1866 and for the field, it clearly emerges that the two groups
present very similar spectroscopic patterns, showing values
consistent within the error bars. 

As far as the light element are concerned (O, Na, Al, Mg), this 
pattern is quite different from what observed in globular cluster stars 
\citep[see e.g. the review by][]{gratton04}  which show two distinctive aspects: 
(i)~the first is that GC stars show a large spread in these light 
elements, indicating inhomogeneous pollution of H burning rich material, and 
(ii)~the second that, because of these effect, the average abundances of GC stars 
are different from those of the field stars with similar metallicity. 
 
We shall emphasize that the chemical abundances of NGC~1866 do not show any 
evidence for these effects: 
{\sl we do not observe appreciable chemical spread within the cluster and the
abundances of NGC~1866 are in very good agreement with those of the LMC field}. 
 
Self-pollution within the cluster, as
originated for example by intermediate AGB stars (e.g.
\citealt{veda}), cannot be completely excluded 
because of  the limited number of stars within
our sample. However we note that in most Galactic GCs
observed with high resolution spectroscopy the percentage of 'polluted' stars is 
significant, at least $\sim$50\% of the entire population
\citep[see e.g.][]{carretta09} and we should expect some clear 
detection within our stars sample. 
As shown in Fig. \ref{anticor} the stars of NGC~1866 well overlap 
the mean locus defined by the giants discussed in \citet{m08}, 
with solar or mild sub-solar [O/Fe] ratios and sub-solar [Na/Fe] 
ratios. This finding, combined with the good agreement between cluster and field stars 
abundance ratios, seems to confirm that all these stars belong 
to the first (unpolluted) generation of the clusters, while there are 
no hints of polluted stars
\footnote{An offset in [O/Fe] between the stars of NGC~1866 and the first generation stars 
of the old LMC and Milky Way GCs is appreciable in Fig.~\ref{anticor}. 
This offset is only due the different chemical evolution of these
clusters: in fact, the first generation stars of the old clusters 
share enhanced [O/Fe] ratios, according to abundances observed in the Halo stars, while 
the stars of NGC~1866 born from a medium enriched by Type Ia SNe, and its 
first generation stars show solar-scaled pattern for the [O/Fe] abundances.
}. 
The lack of anti-correlations in NGC~1866, as far as 
in the intermediate-age, massive LMC clusters, suggests that the younger LMC GCs 
do not undergo the self-enrichment process, following different formation and 
evolution processes with respect to the old stellar clusters (in both Milky Way and the LMC).

Recently, \citet{carretta10} propose to define 
GCs as those stellar clusters where a Na-O anticorrelation is observed. 
This new definition has the appealing advantage to provide an easy boundary 
to separate GCs and other loose stellar systems (as the open clusters).
We stress that this is a {\sl local} definition based only on the Milky Way stellar clusters, 
where there is clear separation in age and mass between open and globular clusters, 
and there is a lack of massive, young stellar clusters (at variance with the LMC).
According to this new definition, NGC~1866 (and all the intermediate-age LMC clusters
so far observed) would 
not be classified as a globular cluster. However, these objects 
appear to be structurally different and more massive than the typical mass 
($<10^4 M_{\odot}$) of the open clusters. Thus, the young populous globular-like 
clusters in the LMC seem to be a class of objects intermediate 
between open clusters and {\sl true} (old) globular clusters.

The main question arising from these findings 
is to understand why these young LMC massive clusters do not 
suffer the self-enrichment process. Previous investigations of old GCs 
show that several parameters (e.g. mass, metallicity, orbital parameters) 
may influence the amount of the self-enrichment process. We note that 
the most metal-rich Galactic clusters (with overall metallicities comparable 
to NGC~1866) are more massive than NGC~1866 by one order of magnitude and thus 
in the Milky Way there are no clusters similar to NGC~1866 
in the mass/metallicity plane.

The chemical homogeneity of NGC~1866 is very important because it 
demonstrates that the chemical inhomogeneities observed in the old GC
stars are peculiar to these objects. NGC~1866 is only a few times less
massive than NGC~6397 and M~4 where inhomogeneities have been observed, so it does
not seem likely that mass alone can be the cause of the differences and
other causes should be invoked, such as, for instance, the fast time
formation of the GC and the (in)homogeneity of the early ISM.  

However, a point to recall is that the young LMC clusters share with several old GCs
the same {\sl present-day} mass but probably not the same {\sl initial} mass. 
In fact, dynamical simulations \citep{dercole08,dercole10} suggest that a large 
fraction of the first stellar generation is lost in the early evolution of the cluster and 
thus the {\sl initial} mass of the cluster was one-two order of magnitude higher 
than the {\sl present-day} mass.
These findings suggest that GCs born with initial mass of the order of $\sim10^{5} M_{\odot}$ 
(similar to the mass of the LMC clusters younger than $\sim$2 Gyr)
are not massive enough to retain their pristine gas and undergo the self-enrichment process.

\subsection{NGC~1866 and evolutive, chemical changes}
 
Since chemical abundance variations can be produced in evolved stars 
by several processes occurring during the stellar evolution, 
as a further step we analyzed the evolutionary status of stars in
our sample, in order to determine whether we could find surface chemical
variations due to events that occurred in their previous evolution.

\begin{figure}
\includegraphics[width=85mm]{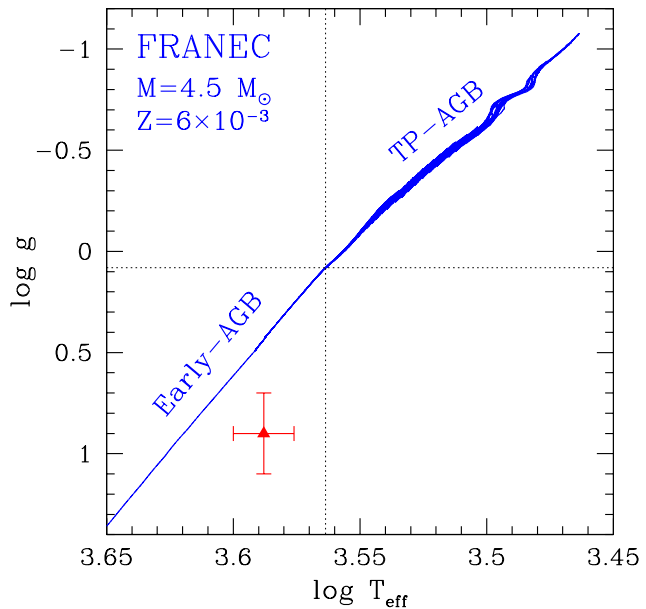}
\caption{Theoretical surface gravity and temperature (blue line)
compared with data relative to \#2981. See text for
details.}
\label{grav}
\end{figure}

\begin{figure}
\includegraphics[width=85mm]{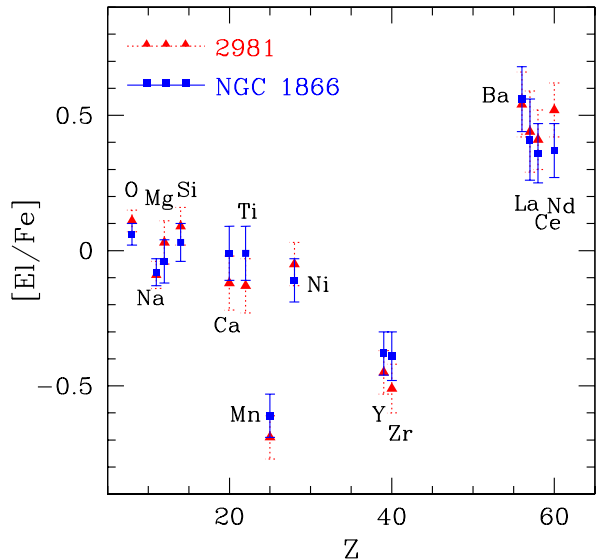}
\caption{Comparison between the spectroscopic values of the
cluster star labelled \#2981 (red triangles) and the median of stars
belonging to the cluster (blue squares).}
\label{agbz}
\end{figure}

The majority of the target stars within our sample lie on their 
RGB and Blue Loop stages and also a few of stars (the brightest 
and reddest ones) belong to the AGB phase.
Therefore, the majority of stars belonging to our
sample have experienced a unique dredge up event, the so-called
First Dredge Up (FDU). Stars belonging to NGC~1866 that evolve
off of their Main Sequence phase have a mass of about $M=4.5
M_\odot$ (according to the evolving mass of the cluster as found
by \citealt{broc03}). Before their first ascent along the Giant
Branch, stellar theory predicts that, in these stars, the FDU
causes a strong depletion of $^{12}$C ($-40$\% $\div -30$\%), a noticeable
enrichment of the surface nitrogen (a factor 2) and a minor
decrease of the oxygen surface abundance. Unfortunately we could
only determine the surface oxygen abundance and, therefore, we
cannot clearly identify the signature of FDU in our stars.
We focus our attention on the most evolved object in our sample 
(the star labelled \#2981) for which we can have a large number of 
elements (due to the large spectral coverage provided from 
UVES). There are other two stars (namely, \#2131 and \#5231) 
that likely belong to the Early-AGB stage, but they are $\sim$200 K 
hotter than \#2981 and some elements cannot be measured due to 
the GIRAFFE spectral coverage. Thus, these two stars are not ideal 
to identify evolutive, chemical changes.\\ 
In order to identify its precise evolutionary phase, we
computed a model of a star with initial mass $M=4.5 M_\odot$ and
$Z=6\times 10^{-3}$ by means of a recent version of the FRANEC
stellar evolutionary code \citep{chi98,sgc06,cri09}. In
Fig.~\ref{grav} we compare the surface gravity and temperature of
the model (blue curve) with data relative to \#2981 (red triangle).
The comparison shows that this star has not yet reached its TP-AGB
phase or, at least, it just suffered for a few TPs. The structure of
an AGB star consists of a partial degenerate C-O core, an
He-shell, an H-shell and a convective envelope. The hydrogen
burning shell, which provides the energy necessary to sustain the
stellar luminosity, is regularly switched off by the growth 
of thermal runaways ({\it Thermal Pulses}, TPs). These
episodes, driven by violent He ignitions within the He buffer
(He-intershell), cause this region to become dynamically unstable
against convection for short periods: once convection quenches off
within the He-intershell, a period of quiet He-burning follows,
during which the convective envelope can penetrate in the
underlying layers (this phenomenon is known as Third Dredge Up,
TDU), carrying to the surface the freshly synthesized carbon and
s-process elements. If the star \#2981 would had already suffered  a
consistent number of TDU episodes,  we would expect
noticeable changes in its s-process surface
abundances\footnote{Note that a previous dredge up event occurring
after the core He-burning (the so-called Second Dredge Up, SDU),
produces minor changes in the CNO surface abundances. However,
variations produced by this event are not easily detectable within
the spectroscopic errors of our sample.}. A comparison between its
spectroscopic data and the median overabundances of the cluster
shows consistent values within error-bars (see Fig.~\ref{agbz}),
therefore supporting the hypothesis that this star is still on its
Early-AGB phase. Unfortunately, spectral lines of some key light
elements (lithium, carbon and nitrogen) are not contained in the observed spectral range.
The abundance of these elements  would provide more
stringent chemical constraints on the evolutionary phase of \#2981,
owing to the occurrence of the already described TDU episodes or
to the presence of other physical processes, such as the Hot
Bottom Burning (HBB) (see, for example, the analysis presented
by \citet{mcs} on their AGB star labeled NGC~1866\#4).

\subsection{ The chemical evolution of the LMC}

\begin{figure}
\includegraphics[width=85mm]{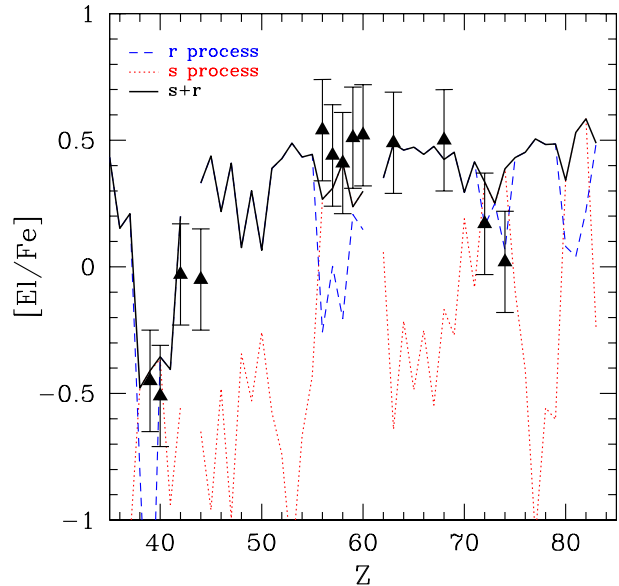}
\caption{Comparison between the
spectroscopic values of \#2981 (black triangle) and the expected
theoretical trend (dark solid line). The single contributions from
the s-process and the r-process are represented by the red dotted
line and the blue dashed line, respectively. See text for
details.} \label{z3m3}
\end{figure}

Our analysis excludes that the spectroscopic patterns
observed in NGC~1866 derive from the evolutionary phase of the observed
stars or from the internal evolution of the cluster: a wider
analysis, which spans over the entire evolutionary history
of the LMC, is therefore necessary. Such an analysis relies on many
physical inputs, the most important being the SFH and the stellar
yields. We just remind that, in the LMC, a rapid chemical enrichment
occurred at a very early epoch, followed by a long period with
reduced star formation and, most recently (about 3 Gyr ago), by
another period of chemical
enrichment \citep[see e.g.][]{bekki}. \\
Concerning the stellar yields, in order to reproduce the heavy
elements ($Z>35$) observed spectroscopic patterns with theoretical
models, we need to hypothesize that two classes of stellar objects
polluted the ISM before the formation of NGC~1866: massive stars,
which synthesized the r-process elements (such as, for example,
europium) and the weak component of the s-process, and AGB stars,
which produced the elements belonging to the main component of the
s-process. In Fig.~\ref{z3m3} we compare our theoretical
expectations with spectroscopic data of \#2981 since, for
this star, we have high resolution spectra and a more
complete element line list at our disposal. 
Note that some of the abundance ratios discussed in the follows are based 
on one only star (see Table~\ref{cl}).
A conservative errorbar of 0.2 dex has been adopted for each element.

\begin{figure}
\includegraphics[width=85mm]{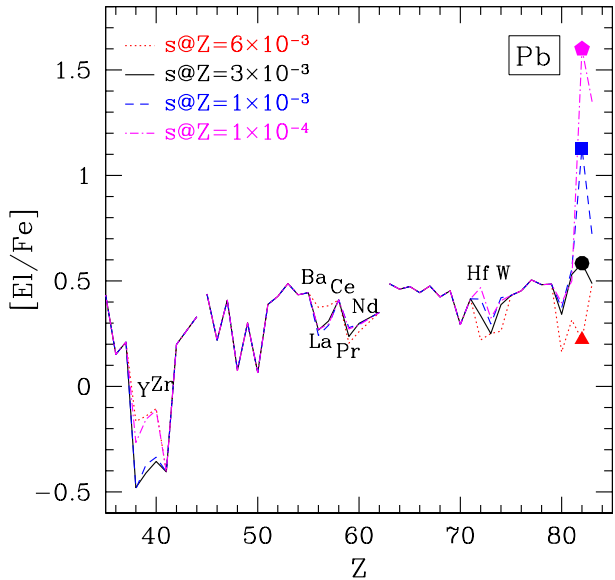}
\caption{Theoretical chemical patterns
obtained with AGB models at different metallicities. See text for
details.} \label{sproc}
\end{figure}

As already discussed, theoretical r-process distributions still
suffer from major uncertainties, such as the identification of the
stellar site or the determination of the precise relative
abundance patterns. For this reason, the r-process contribution to
the solar distribution is usually calculated based on the solar
s-process contribution, following the formula $r=1-s$ (see, e.g.,
\citealt{arla}). Then,
a generic r-process distribution at a fixed metallicity can be
obtained by normalizing the distribution to a single r-only
element (or to an element whose production is almost totally
ascribed to the r-process) and by adopting the solar elemental
ratios for the other elements. We tentatively apply this
procedure, which works well for the Milky Way (see, e.g.,
\citealt{scg08}), to NGC~1866. In order to determine the r-process
enrichment level we focus on europium. We know that about 95\% of
its Galactic abundance can be ascribed to the r-process and we
assume that the same should occur in the Magellanic Clouds. We fix
the europium overabundance to the value of \#2981,
([Eu/Fe]$\sim$0.49\footnote{Note that this value corresponds with
the median europium value calculated over four intermediate-age
LMC clusters of similar metallicity \citep{m08}.}).
Then, we derive the r-process pattern by adopting the elemental
r-process solar percentages tabulated in \citet{biste}. In
Fig.~\ref{z3m3}, the r-process contribution is highlighted  with a
blue dotted line.\\
The s-process contribution has been calculated by means of the
FRANEC code, in which we couple a complete nuclear network (able
to follow in detail the whole s-process nucleosynthesis) directly
to the physical evolution of the model \citep{cri09}. We run, as a
representative mass of AGB pollution, a 2 $M_\odot$ model with
$Z=3\times10^{-3}$ and we hypothesize that the present-day
observed s-process patterns result from the pollution due to a
single generation of low mass AGB stars. This assumption is
justified by the relatively fast chemical evolution of LMC up to
[Fe/H]$\sim$--1 \citep[see e.g.][]{bekki}. Then, we applied a
dilution to the theoretical curve in order to match the cerium
abundance (red dotted curve in Fig.~\ref{z3m3}): this
dilution mimic the fact that the mass lost by AGB stars has been
mixed with s-process free material from which originate
the present-day observed stars.\\
The final theoretical distribution (dark solid curve) results from
the sum of the s-process and the r-process contributions. The
agreement with spectroscopic data is quite good, proving the
validity of our theoretical scheme and validating the assumption
made in the determination of the r-distribution of our sample
(thus possibly evidencing a sort of universality of the
r-process). Unfortunately, the current set of spectroscopic
abundances can not lead us in precisely identifying the
metallicity of AGB population which previously polluted the ISM.
In Fig.~\ref{sproc}, we show different theoretical chemical
patterns (including the r-component) obtained with AGB models of
different metallicities (red dotted line for $Z=6\times10^{-3}$,
dark solid line for $Z=3\times10^{-3}$ (our reference model), blue
dashed line for $Z=1\times10^{-3}$ and magenta dot-dashed line for
$Z=1\times10^{-4}$). Note that, depending on the metallicity,
theoretical models present different enrichment level; before
comparing them, we therefore normalize distributions to the cerium
abundance in order to highlight the relative variations in the
s-process shape. We only highlight the elements, within our
sample, which receive a consistent contribution ($>$50\%) from the
s-process: within error-bars, our spectroscopic data do not permit
us to clearly discriminate between the three distributions. In
order to do that we would need to observe lead, at the termination
of the s-process path, since the abundance of this element is
extremely sensitive to the metallicity. In fact, the lower the
metallicity, the more efficient the Pb production is (see,
e.g., \citealt{biste}): ranging from $Z=1\times10^{-4}$ to
$Z=6\times10^{-3}$ a difference of more than a factor 20 (1.3 dex)
is expected.

Actually, \citet{Rey07} determined the spectroscopic abundances of
elements belonging to the three peaks of the s-process (included
lead\footnote{For this element only an upper limit is available.} in a LMC
post-AGB star (MACHO 47.2496.8). When looking to the relative
distribution, it turns out that the observed path agrees well with our
reference model, whose lead overabundance is comparable to the ones
characterizing the {\it hs} elements. However, more statistics are needed
before claiming any definitive chemical evolutionary theory.

How do our conclusions fit into a more global view of the LMC
chemical evolution? In order to answer to this complex question we
need to compare our data with other LMC samples and to extend our
analysis to abundances of light elements, iron-peak
elements and copper.\\
Concerning heavy elements abundances, stars belonging to LMC
present noticeable differences with respect to their Galactic
counterparts (see Fig.~\ref{yzr} and Fig.~\ref{bala}). In fact,
while in Galactic stars the light elements and heavy elements distributions are nearly
flat (showing values around 0), in LMC they present dichotomic trends. \\
Let us start from the heavy  s-process ({\it hs}) elements. In 2006, \citet{jo06} performed
a spectroscopic analysis on 10 red giants belonging to four old
LMC GCs. Apart from the most metal-poor GC (Hodge 11), which shows
no enhancements at all, in other clusters a mild enhancement of {\it hs}
elements ([hs/Fe]$\sim$0.3 dex) has been found. Similarly, the
study of 27 giants belonging to four intermediate-age LMC GCs by
\citet{mucc08} evidenced a smooth enhancement of heavy elements,
consistent with that found in old LMC GCs. This trend, which also
characterizes metal-poor red giants belonging to dwarf spheroidal
galaxies (dSph) \citep{she03,venn}, can be easily ascribed to a
different SFH of the hosting galaxy. In the LMC, the slower temporal
increase of iron with respect to the Milky Way makes the
contribution from metal-poor AGB stars more important at a given
time or metallicity. Since these objects produce more heavy elements
than light elements, a rise of the heavy elements component has to be expected
(and it is actually observed). Stars belonging to NCG~1866, which
formed only $10^8$ years ago, perfectly match the mild enhancement
observed in others GCs (see Fig.~\ref{bala}). As stressed above, in
order to determine the metallicity of this class of AGB polluters,
the spectroscopic determination of lead is required.\\
Oppositely to {\it hs} elements, light s-process ({\it ls}) elements show a decreasing curve
with respect to Galactic stars at large metallicities. This trend
is fully confirmed by our sample. A similar behaviour has also been
observed in dSph's \citep{venn,she03}: beneath various theoretical
recipes, these authors proposed that these underabundances with
respect to the MW could be ascribed to a reduced contribution from
metal-rich AGB stars or to metallicity dependent yields from SN~II
\citep{tim95}. Both hypotheses are strictly correlated to the
peculiar chemical enrichment that the hosting galaxy experimented
in the past. In LMC, the long gap between the two star formation
bursts has played a fundamental role, melting the contributions
from massive stars and SNIa in a different way with respect to the
MW. A strong reduction in the SFR could have heavily reduced the
contribution from AGB stars of intermediate metallicities, causing
in such a way a decrease of the light elements (note that the yields
of light elements from low mass AGB stars grow with the
metallicity). On the other hand, the behaviour of other elements
efficiently produced by massive stars ($\alpha$ elements, Na, Mn
and Cu) present, at a fixed metallicity, lower overabundances with
respect to the MW (see Figures \ref{ona}, \ref{alfa1} and
\ref{mncu}), suggesting {\sl de facto} a reduced contribution from
massive stars with respect to SN Ia. This statement is however
contrasted by the nearly flat europium distribution observed in
LMC stars ([Eu/Fe]$\sim$0.5) at all metallicities (up to
[Fe/H]$\sim--0.3$)\footnote{We note that a plateau in the
[El/Fe] vs [Fe/H] diagram indicates that the considered
element and iron are produced in equivalent proportions for
different metallicities}. We therefore conclude that a theoretical
analysis based on stellar yields only cannot lead to a clear
explanation for the {\it ls} elements distribution in stars belong to the
LMC. Under this perspective, physical mechanisms involving the
whole LMC structure have to be considered, such for example
dynamical environmental processes \citep{bek} or the presence of
Galactic winds \citep{lanfra}.

\section{Conclusions} \label{concl}

In this paper, we have studied the chemical abundances of 25 stars in the field of the LMC 
star cluster NGC~1866. The accurate analysis and the high efficiency of FLAMES@VLT 
allows us to obtain a set of high quality measurements of the abundances of this 
region of the LMC.  
We emphasize that we do not observe significant element by element abundance 
spread amongst the NGC~1866 stars, and we find that the cluster chemical pattern fits very well 
with the general pattern observed in the LMC field stars.  
We note that this is in stark contrast with 
what is observed with Galactic globular clusters and our result, 
if confirmed on a larger sample of stars, would bring insight to the debate 
of the formation mechanisms for globular clusters in general.

The main observational results are summarized as follows:

1. The average iron abundance of NGC~1866 is [Fe/H]=~--0.43$\pm$0.01 dex ($\sigma$=~0.04 dex).

2. [O/Fe]=~0.07 ($\sigma$=~0.04 dex) and [Na/Fe]=-0.09 ($\sigma$=~0.05 dex )abundance ratios 
appear to be lower than those measured in Galactic stars and the O/Na values are, within the uncertainties, 
very similar between different stars in NGC~1866.

3. The lack of anti-correlations suggests that NGC~1866 does not undergo 
the self-enrichment process at variance with the old GCs in both Milky Way and LMC. 
Similar results have been found in the intermediate-age LMC clusters, suggesting that 
GCs formed with an initial mass of the order of $\sim10^5 M_{\odot}$ are not massive 
enough to retain their pristine gas. Also, other possible effects 
(i.e. a mass/metallicity threshold, inhomogeneity of the early ISM, 
tidal effects due to the interactions with the SMC and the Milky Way) 
cannot be ruled out, playing a role to inhibit the self-enrichment process.

4. $\alpha$-elements in the cluster and in the field stars show a solar-scaled behaviour. 
Also $<\alpha/Fe>$ is measured lower than that found in the Galaxy. 

5. With respect to the Galaxy, a depletion in the abundances of [Mn/Fe] and [Cu/Fe] is found both in 
field and cluster stars. A  value of [Ni/Fe] $\simeq$--0.10 dex is also measured.

6. Abundances of neutron-capture elements are derived: in the case of Y and Zr values lower than 
the solar ones are measured, while  [Ba/Fe], [La/Fe], [Ce/Fe] and [Nd/Fe] ratios appear to be enhanced.
The UVES measurement of a single NGC~1866 star shows a value of [Eu/Fe] $\simeq$~+0.49 dex.

With this observational framework we applied modern stellar evolution theory and nucleosynthesis calculations 
to make three major conclusions. We do caution, however, that our data apply only to a 
single region of the LMC and that abundances of several key elements are lacking, and we hope that 
our work will stimulate further investigations, both observational and theoretical. 
Notwithstanding, the following considerations can be emphasized:

(i)~ The very similar pattern found for the abundances of both field and cluster stars suggests that stars 
belonging to NGC~1866 originate from pollution episodes that occurred before the formation of the cluster. 
Nevertheless, self-enrichment between cluster stars cannot be completely ruled out because of the small number 
of stars.

(ii)~ Surface chemical variations in evolved stars (core He burning and early AGB phases)  due to events 
that occurred in their previous evolution cannot be recognized from data presented in this work. 
Further observations of light elements are recommended to derive more robust constraints.

(iii)~From a relatively simple model we show that the observed abundances of heavy 
elements (Z $>$ 35) can be reproduced by the sum of s-process and r-process contributions as expected by 
pollution mechanisms due to i) massive stars and ii) single generation of low mass AGB stars. 
However, the result obtained
in this work suggest a further theoretical effort to properly understand
the evolution of s-process elements (in particular the {\it ls} ones) in
the context of the LMC chemical evolution. Moreover, precise spectroscopic
measurements of lead are suggested to provide indication on the metallicity
of the low mass AGB stars which could be significant contributors to the
observed abundances of s-process elements in LMC stars.\\

Part of this work has been supported by the Spanish Ministry of
Science and Innovation projects AYA2008-04211-C02-02. 
The authors warmly thank the anonymous referee for his/her 
suggestions in improving the paper and Vanessa Hill for her comments 
and suggestions. A.M. thanks the Observatoire de Meudon, Paris, for its 
hospitality during the early stage of this work.
S.C. thanks Carlos Abia and Roberto Gallino for stimulating discussions


\begin{thebibliography}{}
\bibitem[Alonso et al.(1998)]{alonso98}
Alonso, A., Arribas, S., \& Martinez-Roger, C., 1998, A\&As, 131, 209
\bibitem[Alonso et al.(1999)]{alonso}
Alonso, A., Arribas, S., \& Martinez-Roger, C., 1999, A\&As, 140, 261
\bibitem[Arlandini et al.(1999)]{arla}
Arlandini, C., K\"appeler, F., Wisshak, K., Gallino, R., Lugaro,
M., Busso, M., \& Straniero, O., 1999, ApJ, 525, 886
\bibitem[Bekki \& Chiba(2005)]{bekki}
Bekki, K., \& Chiba, M., 2005, MNRAS, 356, 680
\bibitem[Bekki (2009)]{bek}
Bekki, K., 2009, IAU Symp. 256, 105
\bibitem[Besla et al.(2007)]{besla}
Besla, G. et al., 2007, ApJ, 668, 949
\bibitem[Bisterzo et al.(2009)]{biste}
Bisterzo, S., Gallino, R., Straniero, O., Cristallo, S. \& Kappeler, F. 2010,
MNRAS, 404, 1529
\bibitem[Brocato et al.(2003)]{broc03}
Brocato, E., Castellani, V., Di Carlo, E., Raimondo, G., \& Walker, A. R.,
2003, AJ, 125, 3111
\bibitem[Burris et al.(2000)]{burris}
Burris, D. L., Pilachowski, C. A., Armandroff, T. E.,
Sneden, C., Cowan, J. J., \& Roe, H., 2000, ApJ, 544, 302
\bibitem[Busso, Gallino \& Wasserburg (1999)]{busso}
Busso, M., Gallino, R., \& Wasserburg, G. J.,  1999, ARA\&A, 37, 239
\bibitem[Caffau et al.(2005)]{caffau05}
Caffau, E., Bonifacio, P., Faraggiana, R., Francois, P. Gratton, R. G.,
\& Barbieri, M., 2005, A\&A, 441, 533
\bibitem[Caffau et al.(2008)]{caffau08}
Caffau, E., Ludwig, H.-G., Steffen, M., Ayres, T. R., Bonifacio, P.,
Cayrel, R., Freytag, B., \& Plez, B., 2008, A\&A, 488, 1031
\bibitem[Carpenter(2001)]{carpenter}
Carpenter, J. M., 2001, AJ, 121, 2851
\bibitem[Carretta et al.(2009)]{carretta09}
Carretta, E. et al., 2009, A\&A, 505, 117
\bibitem[Carretta et al.(2010)]{carretta10}
Carretta, E. et al., 2010, A\&A, 516, 55
\bibitem[Castelli \& Kurucz(2003)]{castelli03}
Castelli, F., \& Kurucz, R. L., 2003, in IAU Symposium, Ed. N. Piskunov, W. W. Weiss
\& D. F. Gray, 20P
\bibitem[Cayrel et al.(1999)]{cayrel99}
Cayrel, R., Spite, M., Spite, F., Vangioni-Flam, E., Cass\'e, M. \& Audouze, J., 1999, A\&A, 343, 923
\bibitem[Chieffi et al.(1998)]{chi98}
Chieffi, A., Limongi, M., \& Straniero, O., 1998, ApJ, 502, 737
\bibitem[Cole et al.(2005)]{cole}
Cole, A. A., Tolstoy, E., Gallagher, J. S., III, \& Smecker-Hane, T. A., 2005, AJ,
129, 1465
\bibitem[Colucci, Bernstein and McWilliam(2010)]{colucci}
Colucci, J.E., Bernstein, R. A., \& McWilliam, A., 2010, arXiv:1009.4195v1
\bibitem[Cristallo et al.(2009)]{cri09}
Cristallo, S., Straniero, O., Gallino, R., Piersanti, L.,
Dom\'inguez, I. \& Lederer, M.T., 2009, ApJ, 696, 797
\bibitem[Cunha et al.(2002)]{cunha}
Cunha, K., Smith, V. V., Suntzeff, N. B., Norris, J. E.,
Da Costa, G. S., \& Plez, B., 2002, AJ, 124, 379
\bibitem[Den Hartog et al.(2003)]{den}
Den Hartog, E. A., Lawler, J. E., Sneden, C., \& Cowan, J.J., 2003, ApJS,
148, 543
\bibitem[D'Ercole et al.(2008)]{dercole08}
D'Ercole, A., Vesperini, E., D'Antona, F., McMillan, S. L. W., \& Recchi, S., 2008, 
MNRAS, 391, 825
\bibitem[D'Ercole et al.(2010)]{dercole10}
D'Ercole, A., D'Antona, F., Ventura, P., Vesperini, E., \& McMillan, S. L. W., 2010, 
MNRAS, 407, 854
\bibitem[Edvardsson et al.(1993)]{edv}
Edvardsson, B., Andersen, J., Gustafsson, B., Lambert, D. L., Nissen, P. E., \&
 Tomkin, J., 1993, A\&A, 275, 101
\bibitem[Fulbright(2000)]{ful}
Fulbright, J. P., 2000, AJ, 120, 1841
\bibitem[Gallino et al.(1998)]{ga98}
Gallino, R., Arlandini, C., Busso, M., Lugaro, M., Travaglio, C,
Straniero, O., Chieffi, A. \& Limongi, M., 1998,ApJ, 497, 388
\bibitem[Gratton et al.(1999)]{gr99}
Gratton, R. G., Carretta, E., Eriksson, K., \& Gustafsson, B., 1999, A\&A, 350, 955
\bibitem[Gratton et al.(2003)]{gr03}
Gratton, R. G., Carretta, E., Claudi, R., Lucatello, S., \& Barbieri, M., 2003, A\&A,
404, 187
\bibitem[Gratton, Sneden \& Carretta(2004)]{gratton04}
Gratton, R. G., Sneden, C. \& Carretta, E., 2004, ARA\&A, 42, 385
\bibitem[Grevesse \& Sauval(1998)]{gs98}
Grevesse, N., \& Sauval, A. J., 1998, SSRv, 85, 161
\bibitem[James(1998)]{james}
James, F., 1998, MINUIT, Reference Manual, Version 94.1, CERN, Geneva,
Switzerland
\bibitem[Johansson et al.(2003)]{j03}
Johansson, S, Litzen, U., Lundberg, H., \& Zhang, Z., 2003, ApJ, 584, 107L
\bibitem[Johnson et al.(2006)]{jo06}
Johnson, A.J., Ivans, I.I., \& Stetson, P.B., 2006, ApJ, 640, 801
\bibitem[Harris \& Zaritsky(2009)]{harris09}
Harris, J., \& Zaritsky, D., 2009, AJ, 138, 1243
\bibitem[Hill et al.(2000)]{hill00}
Hill, V., Francois, P., Spite, M., Primas, F., \& Spite, F., 2000, A\&As, 364, 19
\bibitem[Hodge (1960)]{hdg60}
Hodge, P. W.,  1960, ApJ , 131, 351
\bibitem[Hodge (1961)]{hdg61}
Hodge, P. W.,  1961, ApJ , 133, 413
\bibitem[Kratz et al.(2007)]{kra}
Kratz, K.-L., Farouqi, K., Pfeiffer, B., Truran, J.W., Sneden, C., Cowan, J.J., 2007, ApJ, 662, 39
\bibitem[Kurucz(1993a)]{kur93a}
Kurucz, R. L., 1993a, ATLAS9 Stellar Atmosphere Programs and 2 km/s grid.
Kurucz CD-ROM No. 13. Cambridge, Mass,: Smithsonian Astrophysical Observatory,
1993., 13
\bibitem[Kurucz(1993b)]{kur93b}
Kurucz, R. L., 1993b, SYNTHE Spectral Synthesis Programs and Line Data.
Kurucz CD-ROM No. 18. Cambridge, Mass,: Smithsonian Astrophysical Observatory,
1993., 18
\bibitem[Lanfranchi et al.(2008)]{lanfra}
Lanfranchi, G.A., Matteucci, F., \& Cescutti, G., 2008, A\&A, 481,
635
\bibitem[Lawler et al.(2001a)]{law1}
Lawler, J. E., Wickliffe, M. E., den Hartog, E. A., \& Sneden, C., 2001, ApJ, 563, 1075
\bibitem[Lawler et al.(2001b)]{law2}
Lawler, J. E., Bonvallet, G., \& Sneden, C., 2001, ApJ, 556, 452
\bibitem[Lodders, Palme \& Gail(2009)]{lodders}
Lodders, K., Palme, H., \& Gail, H.-P., 2009, arXiv0901.1149L
\bibitem[Magain(1984)]{magain}
Magain, P. 1984, A\&A, 134, 189
\bibitem[Matteucci \& Brocato (1990)]{mb90}
Matteucci, F., \& Brocato, E., 1990, ApJ, 365, 539
\bibitem[McSaveney et al.(2007)]{mcs}
McSaveney, J.A., Wood, P.R., Scholz, M., Lattanzio, J.C., \& Hinkle,
K.H., 2007, MNRAS, 378, 1089
\bibitem[Mucciarelli et al.(2006)]{m06a}
Mucciarelli, A., Origlia, L., Ferraro, F. R., Testa, V., \& Maraston, C., 2006,
ApJ, 646, 939
\bibitem[Mucciarelli et al.(2008a)]{mucc08}
Mucciarelli, A., Caffau, E., Freytag, B., Ludwig, H.-G., \& Bonifacio, P., 2008,
A\&A, 484, 841
\bibitem[Mucciarelli et al.(2008b)]{m08}
Mucciarelli, A., Carretta, E., Origlia, L. \& Ferraro, F. R., 2008,
AJ, 136, 375
\bibitem[Mucciarelli et al.(2009)]{m09}
Mucciarelli, A., Origlia, L., Ferraro, F. R., \& Pancino, E., 2009, ApJ, 695, 134L
\bibitem[Mucciarelli et al.(2010)]{m10}
Mucciarelli, A., Origlia, L.\&  Ferraro, F. R.2010, ApJ, 717, 277
\bibitem[Musella et al.(2006)]{musella06}
Musella, I., Ripepi, V., Brocato, E., Castellani, V., Caputo, F., 
Del Principe, M., Marconi, M., Piersimoni, A. M., Raimondo, G., 
Stetson, P. B., \& Walker, A. R., 2006, MemSaIt, 77, 291
\bibitem[Pasquini et al.(2002)]{pasquini02}
Pasquini, L. et al., 2002, Messenger, 110, 1
\bibitem[Pompeia et al.(2005)]{pomp5}
Pompeia, L., Hill V. \& Spite, M., 2005, NuPhA, 758, 242
\bibitem[Pompeia et al.(2008)]{pomp}
Pompeia, L., et al., 2008, A\&A, 480, 379
\bibitem[Prochaska et al.(2000)]{prochaska}
Prochaska, J. X., Naumov, S. O., Carney, B. W., McWilliam, A., \& Wolfe, A.,
2000, AJ, 120, 2513
\bibitem[Qian \& Wasserburg (2007)]{qw}
Qian, Y.-Z., \& Wasserburg, G.J., 2007, PhR, 442, 237
\bibitem[Reddy et al.(2003)]{reddy}
Reddy, B. E., Tomkin, J., Lambert, D. L., \& Allende Prieto, C., 2003, MNRAS, 340, 304
\bibitem[Reddy et al.(2006)]{reddy06}
Reddy, B. E., Lambert, D. L., \& Allende Prieto, C., 2006, MNRAS, 367, 1329
\bibitem[Reyniers et al.(2007)]{Rey07}
M. Reyniers, C. Abia, H. Van Winckel, T. Lloyd Evans, L. Decin, K.
Eriksson, \& K. R. Pollard, 2007, A\&A, 461, 641
\bibitem[Rieke \& Lebofsky(1985)]{rl85}
Rieke, G. H., \& Lebofsky, M. J., 1985, ApJ, 288, 618
\bibitem[Sbordone et al.(2004)]{sbordone04}
Sbordone, L., Bonifacio, P., Castelli, F., \& Kurucz, R. L.,
2004, MemSaIt, 5, 93
\bibitem[Shetrone et al.(2003)]{she03}
Shetrone, M., Venn, K.A., Tolstoy, E., Primas, F., Hill, V., \&
Kaufer, A., 2003, AJ, 125, 684
\bibitem[Smith et al.(2002)]{smith}
Smith, V.V., et al., 2002, AJ, 124, 3241
\bibitem[Sneden et al.(2008)]{scg08}
Sneden, C., Cowan, J.J., \& Gallino, R., 2008, ARA\&A, 46, 241
\bibitem[Staveley-Smith et al.(2003)]{stav}
Staveley-Smith, L., Kim, S., Calabretta, M. R., Haynes, R. F., \&
Kesteven, M. J., 2003, MNRAS, 339,87
\bibitem[Straniero et al.(2006)]{sgc06}
Straniero, O., Gallino, R., \& Cristallo, S., 2006, Nucl. Phys. A,
777,311
\bibitem[Storey \& Zeippen(2000)]{storey}
Storey, P. J., \& Zeippen, C. J., 2000, MNRAS, 312, 813
\bibitem[Timmes et al.(1995)]{tim95}
Timmes, F., Woosley, S.E., \& Weaver, T.A., 1995, ApJS, 98, 617
\bibitem[Tolstoy, Hill \& Tosi (2009)]{tht09}
Tolstoy, E., Hill, V., \& Tosi, M., 2009, ARA\&A, 47, 371
\bibitem[van den Bergh \& Hagen(1968)]{vdb68}
van den Bergh, S. \& Hagen, G. L., 1968, AJ, 73, 569
\bibitem[van den Bergh \& de Boer(1984)]{vdbdb84}
van den Bergh, S. \& de Boer, K. D., 1984, Structure and evolution of the Magellanic Clouds, 
Proceedings of the 108th IAU Symposium Dordrecht: Reidel 1984
\bibitem[Venn et al.(2004)]{venn}
Venn, K. A., Irwin, M. Shetrone, M. D., Tout, C. A., Hill, V.,
\& Tolstoy, E., 2004, AJ, 128, 1177
\bibitem[Ventura \& D'Antona (2009)]{veda}
Ventura, P., \& D'Antona, F., 2009, A\&A, 499, 835
\bibitem[Wahlgren(2005)]{wal}
Wahlgren, G. M., 2005, MemSaIt Suppl.,
8, 108
\bibitem[Walker et al.(2001)]{walker}
Walker, A. R., Raimondo, G., Di Carlo, E., Brocato, E.,
Castellani, V., \& Hill, V., 2001, ApJ, 560, 139L

\end{thebibliography}
\end{document}